\documentclass[traditabstract]{aa} 
\usepackage{amsmath}
\usepackage{graphicx}
\usepackage{siunitx}
\usepackage{txfonts}
\usepackage{graphicx}
\usepackage{caption}
\usepackage{natbib}
\usepackage{subfigure} 
\usepackage{hyperref}
\usepackage[utf8]{inputenc}
\usepackage{placeins}
\usepackage{scrextend}
\usepackage{xcolor,colortbl}
\usepackage{rotating} 
\usepackage{multirow} 
\usepackage{booktabs} 
\usepackage{array}
\usepackage{soul}
\newcolumntype{P}[1]{>{\centering\arraybackslash}p{#1}}

\newcommand{\bd}{\begin{displaymath}}
\newcommand{\ed}{\end{displaymath}}
\newcommand{\be}{\begin{equation}}
\newcommand{\ee}{\end{equation}}
\newcommand{\beaa}{\begin{eqnarray*}}
\newcommand{\eeaa}{\end{eqnarray*}}
\newcommand{\bea}{\begin{eqnarray}}
\newcommand{\eea}{\end{eqnarray}}

\newcommand{\erg}{\mathrm{erg}}

\DeclareSIUnit\parsec{pc}
\DeclareSIUnit\lightyear{ly}
\DeclareSIUnit\year{yr}
\DeclareSIUnit\erg{erg}
\DeclareSIUnit\ster{ster}
\DeclareSIUnit\arcsec{arcsec}
\DeclareSIUnit\rad{rad}
\DeclareSIUnit\mag{mag}

\usepackage{color} 
\definecolor{gruen}{cmyk}{0.35,0.01,0.80,0.1}

\definecolor{blue}{rgb}{0.0,0.0,1.0}
\definecolor{magenta}{rgb}{1.0,0.0,1.0}
\definecolor{orange}{rgb}{1.0,0.5,0.0}

\def\kmsMpc {\rm km\,s^{-1}\,Mpc^{-1}}

\begin{document}

   \title{HOLISMOKES - XIV. Time-delay and differential dust extinction determination with lensed type II supernova color curves}
  \titlerunning{HOLISMOKES - XIV. Time-delay and differential dust extinction determination with LSN II color curves}

   \author{J. Grupa\inst{1,2,3}
   			\and
   			S. Taubenberger\inst{1,2}
                          \and 
          S. H. Suyu\inst{2,1}
   		 \and 
   		 S. Huber \inst{1,2}
                  \and 
             C. Vogl\inst{1,2,3}
             \and 
             D. Sluse\inst{4}
          }

   \institute{Max-Planck-Institut f\"ur Astrophysik, Karl-Schwarzschild Str. 1, 85748 Garching, Germany\\
              \email{jana@MPA-Garching.MPG.DE}
         \and 
           Technische Universit\"at M\"unchen, TUM School of Natural Sciences, Physics Department,  James-Franck-Stra\ss{}e~1, 85748 Garching, Germany
			\and 
           Exzellenzcluster ORIGINS, Boltzmannstr. 2, 85748 Garching, Germany
\and
           STAR Institute, Quartier Agora – Allée du six Août, 19c, 4000 Liège, Belgium}


 
  \abstract
  {The Hubble tension is one of the most relevant unsolved problems in cosmology today. Strongly gravitationally lensed transient objects, such as strongly lensed supernovae, are an independent and competitive probe that can be used to determine the Hubble constant. In this context, the time delay between different images of lensed supernovae is a key ingredient.
We present a method, to retrieve time delays and the amount of differential dust extinction between multiple images of lensed type IIP supernovae through their color curves, which display a kink in the time evolution.
With multiple realistic mock color curves based on an observed unlensed supernova from the Carnegie Supernova Project, we demonstrate that we can retrieve the time delay with uncertainties of $\pm 1.0$ days for light curves with 2-day cadence and 35\% missing data due to weather losses. The differential dust extinction is more susceptible to uncertainties, because it depends on imposing the correct extinction law.
Further we also investigate the kink structure in the color curves for different rest-frame wavelength bands, particularly rest-frame UV light curves from SWIFT, finding sufficiently strong kinks for our method to work for typical lensed SN redshifts that would redshift the kink feature to optical wavelengths.
With the upcoming Rubin Observatory Legacy Survey of Space and Time, hundreds of strongly lensed supernovae will be detected and our new method for lensed SN IIP is readily applicable to provide delays.}


   \maketitle

\section{Introduction}
\label{sec:intro}

The determination of the Hubble constant $H_{0}$ has been one of the most relevant goals in cosmology in the past decades. It defines the universe's current expansion rate and, therefore, sets the scale and age of the universe today.
Several methods to determine $H_{0}$ have already been applied \citep{Valentino2021}, but there is still a 
$>$4$\sigma$ tension \citep{Verde2019} between cosmic microwave background measurements by the \cite{Planck2020} and distance ladder measurements from the SH0ES (Supernova $H_{0}$ for the Equation of State) program \citep{Riess2022, Riess2024,Li2024,Anand2024} as well as other late universe measurements. 
The most recent results from distance ladder calibrations with the Tip of the Red Giant Branch (TRGB) from \cite{Freedman2021}\citep[see also][]{Freedman2019,Freedman2020} are in agreement within 1.3$\sigma$ with the \cite{Planck2020} and at a 1.6$\sigma$ level with \cite{Riess2022}. Another usage of the TRGB from \cite{Soltis2021} is in better agreement with \cite{Riess2022}.

The Hubble tension hints at either new unknown physics beyond our current model of the universe or at problems within the various $H_{0}$ measurements or distance calibrations. Therefore, we aim for determining $H_{0}$ without distance calibrations. Three such methods are using radiative transfer modeling of type II supernovae (SNe II) based on the tailored-expanding-photosphere method \citep[e.g.][]{Vogl2020,Dessart2009,Dessart2008,Dessart2006}, gravitational wave sources \citep[e.g.][]{Palmese2024,Mukherjee2021,Gayathri2021}, or masers \citep{Pesce2020}.
A further direct method are time-delay measurements of distant gravitationally lensed transient objects as first proposed by \cite{Refsdal1964}.
Such objects can be quasars or SNe, which show up as multiple images if they are strongly gravitationally lensed. The light rays of these images have shorter or longer light paths as they pass through different parts of the gravitational potential, leading to a time difference between the image appearances. As transient objects show features like peaks and kinks in light and color curves, we can infer the time delay between images of such objects. Together with models of the gravitational lensing system and reconstructions of the mass perturbations along the line of sight, we can calculate $H_{0}$.

This technique has already been used by the H0LiCOW ($H_{0}$ Lenses in COSMOGRAIL’s Wellspring) program \citep{Suyu2017,Birrer2018,Sluse2019,Chen2019,Wong2019,Rusu2019} collaborating with the COSmological MOnitoring of GRAvItational Lenses \citep[COSMOGRAIL;][]{Eigenbrod2005, Courbin2017, Bonvin2018} and the Strong lensing at High Angular Resolution Program \citep[SHARP;][]{Chen2019}. Combining the time-delay measurements of six quasars, the Hubble constant is determined as $H_{0} = 73.3 ^{+1.7} _{-1.8} \ \mathrm{km}\mathrm{s^{-1}}\mathrm{Mpc^{-1}}$ assuming a flat $\Lambda$CDM cosmology and physically motivated lens mass distributions \citep{Wong2019}. This value is in agreement with late universe measurements. The STRong-lensing Insights into the Dark Energy Survey (STRIDES) collaboration \citep{Treu2018} analyzed a seventh lensed quasar system \citep{Shajib2020} which agrees well with \cite{Wong2019}.
Systematic effects are further tested and more lensed quasars are studied by the Time-Delay COSMOgraphy (TDCOSMO) organization \citep{Millon2020a,Millon2020, Gilman2020, Birrer2020}.

Originally \cite{Refsdal1964} suggested using lensed supernovae (LSNe) as the transient source for time-delay cosmography. The first spatially resolved LSN called ``SN Refsdal", which was of type II, was detected 50 years later \citep{Kelly2015,Kelly2016a,Kelly2016b}. Since then a few more LSNe have been observed \citep{Goobar2017, Chen2022, Goobar2022}. 
The most recently detected LSNe are ``SN H0pe" \citep{Frye2023,Frye2024}, which has been confirmed to be of type Ia, and ``SN Encore" \citep{Pierel2024}, which appeared in a lensing system that has previously hosted another lensed SN, most likely of type Ia \citep{Rodney2021}.
With SN Refsdal and SN H0pe, the first time-delay cosmography measurements have been performed \citep{Kelly2023,Grillo2024,Liu2024,Pascale2024,Pierel2024}.
Lensed supernovae are not only interesting in terms of cosmography, but also give insights into their explosion mechanisms, as the modeling of the gravitational potential can predict the temporal and spatial location of future images, which gives the opportunity to study them from very early on.

The performance of LSNe Ia for time-delay cosmography has been studied in several papers \citep[e.g.,][]{Dobler2006, Goldstein2018, Huber2019,Huber2022,Suyu2020,Huber2024, Arendse2024}. SNe II have not been given that much attention, even though they are the most abundant type of SNe by volume \citep{Li2011} and we expect to detect hundreds of LSNe II with the upcoming Rubin Observatory Legacy Survey of Space and Time \citep[LSST;][]{LSSTScienceCollaboration2009} \citep{Wojtak2019, Goldstein2018, Goldstein2016}. This is the reason we focus on this type in our work, in particular SNe IIP.
A major concern for the correct time-delay determination are microlensing effects stemming from small and compact masses within the lens galaxy, such as stars. In paper HOLISMOKES V \citep{Bayer2021}, we found that in the early plateau phase of SNe IIP during at least the first 30 days after explosion, the specific intensity profiles are achromatic as in the case of SNe Ia as well \citep{Goldstein2018,Huber2019,Huber2021}. Therefore, we can neglect microlensing effects on the color curves during these phases, as the additional magnification of microlensing cancels out in the color curve. 

\begin{figure}[hbt!]
\centering
\includegraphics[width=0.49\textwidth]{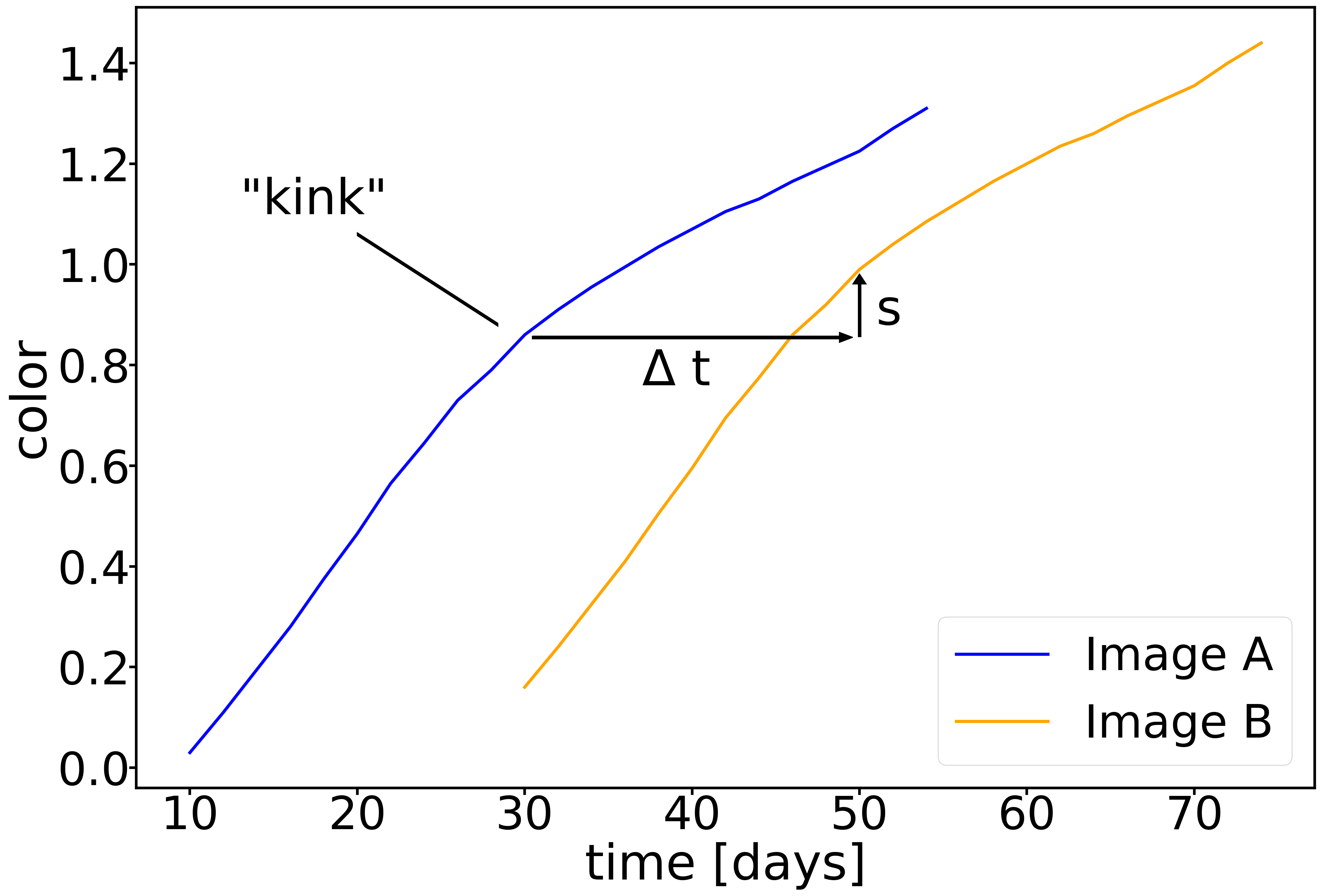}
\caption{\label{color_curve_scheme}Exemplary noiseless LSN IIP color curves of two lensing images A and B showing a time shift $\Delta t$ and a shift in color $s$ due to extinction.}
\end{figure}

While we focused on getting time delays from spectra in HOLISMOKES V, this work now focuses on the determination of a time delay, based on the presence of a small kink in the color curves during the plateau phase \citep{deJaeger2018}. As photometric follow-up is much less expensive than a spectroscopic one, this method should be more easily applicable for studying a large sample of objects. 
An arbitrary example of SN IIP color curves of two lensing images is shown in Fig. \ref{color_curve_scheme}, illustrating the time delay $\Delta t$ and a shift in color $s$, caused by the different extinction for the two images A and B.
We use SN data from the Carnegie Supernova Project (CSP) \citep{CSP} to construct mock color curves and determine time delays and extinction parameters \citep{Cardelli1989}
with the help of Monte Carlo Markov Chain (MCMC) sampling \citep{Dunkley2005}.

Following the findings in \cite{Suyu2020}, we would need cosmology grade time-delay measurements with uncertainties $\lesssim$ 5\% of about 20 spatially resolved LSNe Ia, in order to determine $H_{0}$ with an uncertainty of $\sim$ 1\%, under the realistic assumption of $\lesssim$ 3\% lens mass modelling uncertainties. As we expect to find more LSNe II than LSNe Ia, including also other types than just SNe IIP, during the era of LSST and we find in HOLISMOKES V that LSNe II are less susceptible to microlensing resulting in less biased time delays, even systems with time-delay precision $\lesssim$ 10\% will help to achieve the desired precision on the Hubble constant. 

We structure our work as follows:
In Sect. \ref{sec: type II supernova color curves} we go into detail on dust extinction and the creation of our mock SN IIP color curves. We present our MCMC sampling in Sect. \ref{sec:MCMC} and the results in Sect. \ref{sec:results} applying it to the rest-frame SN mock color curves. After that we look into the impact of redshift in Sect. \ref{sec: K_corr} 
We finish this work with a discussion and conclusion in Sect. \ref{sec: Discussion_Conclusion}.

\section{Type II supernova color curves}
\label{sec: type II supernova color curves}


\cite{deJaeger2018} found that SN II color curves show a characteristic change in their slope within the plateau phase, which lasts from (rest-frame) days of $\sim$10 to $\sim$100 after explosion. Before the kink around day 35 (rest-frame) after explosion, the color becomes redder more rapidly than afterwards.
This ``kink" (as shown in Fig. \ref{color_curve_scheme}) enables us to do cosmography as well as determine extinction parameters.

In this section, we detail the creation of the mock data used in this work. In Sect. \ref{sec: dust extinction} we introduce the extinction laws by \cite{Cardelli1989} referred to as Cardelli laws. Sect. \ref{sec: light curves} shows the CSP light curves that we use as archetypical SN IIP light curves. We fit these light curves with orthogonal polynomial functions in Sect. \ref{sec:light_curve_model} to create the mock data in \ref{sec:color_curves}.

\subsection{Dust extinction and extinction parameters}
\label{sec: dust extinction}

The multiple images of a lensed supernova are observed through different regions of the lensing galaxy. For that reason, it is necessary to emulate the impact of dust on the colors of the lensed images. For this experiment, we use the Cardelli extinction laws, which provides a flexible model of the extinction in galaxies \citep{Cardelli1988,Cardelli1989}.
The Cardelli laws consider dust absorption and reemission at longer wavelengths as well as light scattering away from the line of sight. Back scattering of light of surrounding stars into the line of sight is not included as our source is far away.
We further ignore extinction occurring in the Milky Way, as we assume it to be the same for the different lensed images owing to the small angular separation.
We remark that the Cardelli laws for mock data creation can be exchanged by any other flexible model of extinction in galaxies, such as the more recent extinction laws by \cite{Gordon2023} without impacting the retrieval of the time delay.

We denote the intrinsic magnitude as $m_{\lambda,0}$ and the observed magnitude as $m_{\lambda}$ at the wavelength $\lambda$. The wavelength-dependent extinction coefficient is defined as:
\begin{equation}
A_{\lambda} = m_{\lambda} - m_{\lambda,0}.
\end{equation}
Extinction laws are typically defined as $A_{\lambda} / A_{V}$, for which $A_{V}$ is a normalization defined in the optical at $\lambda \approx 5500 \ \AA$. They can also be expressed via the total-to-selective-extinction ratio \citep{Cardelli1989}:
\begin{equation}
R_{V} =  \frac{A_{V}}{E(B - V)} = \frac{A_{V}}{(A_{B} - A_{V})},
\end{equation}
depending on the color excess:
\begin{equation}
E(B-V) \equiv A_{B} - A_{V}.
\end{equation}

The Cardelli laws \citep{Cardelli1988,Cardelli1989} are generally defined depending only on $R_{V}$:
\begin{equation} \label{eq: Cardelli}
\frac{A_{\lambda}}{A_{V}} = a(x) + \frac{b(x)}{R_{V}}.
\end{equation}
The parametric functions $a(x)$ and $b(x)$, where $x = \lambda^{-1}$, are determined analytically from spline fits to measured data in the three wavelength regimes, infrared (IR), optical, and ultraviolet (UV) \citep{Cardelli1988a,Cardelli1988b}.
The extinction impacts the color curve via a magnitude shift $s$ (see Fig. \ref{color_curve_scheme}), which is correlated for the case of multiple color curves.

\subsection{Observed light curves}
\label{sec: light curves}

Until today, only two spatially resolved lensed SNe II have been observed. Therefore, we have to create mock color curves based on measured light curves of unlensed SNe II.
We base our mock color curves on observed light curves from CSP.
The data used in this work from CSP (Anderson et al. in prep.) are rest-frame $u,g,$ and $r$ band light curves (see \citealp{CSP_filters} for the definition of the passbands) for SN2005J, which are shown in Fig. \ref{u_g_fit}. We treat these filters as universally applicable to other facilities, which might be used in the future for follow up observations in the LSST era. The corresponding rest-frame color curves $u-g$ and $u-r$, which are used later on in the study, show a strong kink mainly because of the larger brightness decrease in $u$ band compared to the $g$ and $r$ bands. The impact of redshift will be discussed later in Sect. \ref{K_corr}.

\subsection{Light curve model}
\label{sec:light_curve_model}

We first derive a parametrized model of the light curves, which we can use to generate mock light curves and color curves of LSNe IIP. We use orthogonal polynomials as the fitting function $C(\vec{\eta},t)$ that is applied to the observed light curves, where $\vec{\eta}$ is the vector of fitting parameters. In addition to the orthogonal polynomial fitting procedure, we test spline fitting in Appendix \ref{App:diff_model} to test the sensitivity of our result to the  choice of model.

For the fitting with orthogonal polynomial functions, we consider a linear combination of Legendre polynomials $P_{n}$ as the functional form $C(\vec{\eta}_{\mathrm{model}},t)$ on an interval $[t_{\mathrm{min}},t_{\mathrm{max}}]$ up to the order of $n=$ 6:
\begin{equation}\label{eq:5}
C(\vec{\eta}_{\mathrm{model}},t) = c_{1} P_{0} + c_{2} P_{1} + c_{3} P_{2} + ... + c_{n+1} P_{n},
\end{equation}
where $P_{n}$ is defined as:
\begin{equation}
P_{n} = \frac{1}{2^{n}n!}\frac{\mathrm{d}^{n}}{\mathrm{d}u^{n}}(u^{2}-1)^{n},
\end{equation}
with substitution $u = (2t-t_{\mathrm{min}}-t_{\mathrm{max}}) / (t_{\mathrm{max}}-t_{\mathrm{min}})$ to ensure that the Legendre polynomials are orthogonal on an interval $t = [t_{\mathrm{min}},t_{\mathrm{max}}]$ (with $u = [-1,1]$) and $P_{0} = 1$. The orthogonality interval $[t_{\mathrm{min}},t_{\mathrm{max}}]$ depends on the time range covered by the color curve of the lensed image observed for the longest period. In our case we can take either the first or the second image time range, as both are of equal length. For a real measurement this will be most likely the second image, as its light curve can be followed and gathered at an earlier time compared to the first image. We choose the highest order as $n$ = 6 as it is flexible enough.

\begin{figure}[hbt!]
\centering
\includegraphics[width=0.49\textwidth]{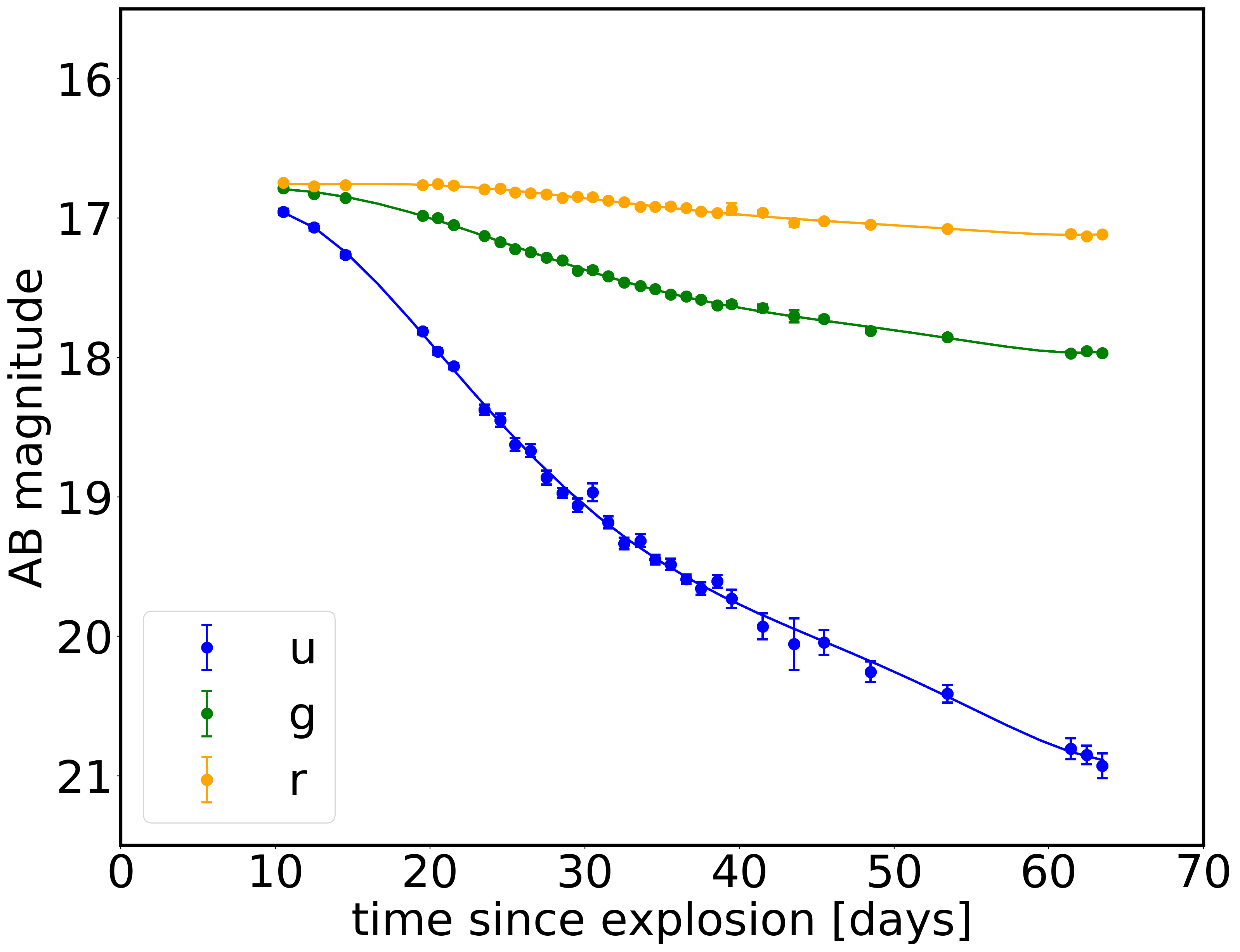}
\caption{\label{u_g_fit} $u,\ g,$ and $r$ light curves of SN2005J observed by the CSP (Anderson et al. in prep.) including examples of orthogonal polynomial fits (solid lines) to the $u$, $g$ and $r$ band light curves.}
\end{figure}

The parameter vector is defined as $\vec{\eta}_{\mathrm{mock}} = (c_{1},c_{2},c_{3},c_{4},c_{5},c_{6},c_{7})_{\mathrm{light \ curve}}$.
For the mock data creation, we fit the measured CSP data several times with $C(\vec{\eta}_{\mathrm{mock}},t)$ to ensure reaching the global minimum and select the fit with the lowest $\chi^{2}$.
An example of a light curve fit with up to 6th order orthogonal polynomials to the light curves $u$, $g$, and $r$ can be seen in Fig. \ref{u_g_fit}.

\subsection{Mock data creation}
\label{sec:color_curves}

To generate mock light curves, we assume systems detected by LSST \citep{Oguri2010}, but followed up at high cadence at a different facility. The parametric SN model as introduced in Sect. \ref{sec:light_curve_model} is evaluated with an artificial two-day cadence, where each simulated data point can vary around the perfect two-day cadence to simulate realistic observing conditions adding a value randomly picked from an uniform distribution [$-0.1$,0.1].

Uncertainties are added following the procedure of \cite{Huber2022} and \cite{LSSTScienceCollaboration2009} assuming a follow up facility with a 2 m main mirror.
Specifically, at each observational time $t_{i}$ and for each observational band, a random number $g_{i}$ is drawn from a Gaussian with zero mean and standard deviation $q_{i}$. These numbers are then added to the value determined from the fitted function. The value of $q_{i}$ is set by the signal-to-noise ratio (S/N) of the observed magnitude, which depends on $C(\vec{\eta}_{\mathrm{mock}},t_{i})$ and the limiting magnitude, which we adopt as one magnitude deeper than the LSST-like $5\sigma$ depth \citep{Huber2019}. If $C(\vec{\eta}_{\mathrm{mock}},t_{i})$ is more than two magnitudes fainter than the $5 \sigma$ depth, the data point is not generated.
Following this approach, we create the mock light curves for the first strongly lensed image A. The mock light curves of the second image B are created similarly, but adding a time and magnitude shift from the filter-dependent extinction and evaluating the model at time steps coinciding with the ones from image A, to ensure measurements are taken at the same time for both A and B.
For each image, we subtract the mock light curves in different filters to create the mock color curves $\vec{d}_{\mathrm{A}}$ and $\vec{d}_{\mathrm{B}}$. The color curve uncertainties $\vec{\sigma}_{\mathrm{A}}$ and $\vec{\sigma}_{\mathrm{B}}$ are calculated by adding the light curve uncertainties in quadrature.

For the main study, we investigate three different scenarios in the mock data creation with SN2005J, which are shown in Table \ref{table_cases}:
\begin{enumerate}
		\item High S/N assuming the uncertainties given by unlensed low redshift observations (Anderson et al. in prep.)
		\item Low S/N assuming the V band light curve peak of the first image to be at 22 mag and for the second image to be at 23 mag
		\item Low S/N assuming the V band light curve peak of the first image to be at 22 mag and for the second image to be at 23, mag with additionally 35\% of the data missing to simulate bad weather conditions and other incidents where observations are not possible
\end{enumerate}

\begin{table*}[htb]
\centering
\renewcommand*{\arraystretch}{1.1}
\begin{tabular}{|m{2.4cm}|m{4.9cm}|m{4.65cm}|m{4.65cm}|}
\hline 
\centering Scenario & \centering 1: high S/N & \centering 2: low S/N & $\ \ \ \ $ 3: low S/N with 35\% data loss \\ 
\hline
\centering \# of mocks & \centering 100 & \centering 500 & $\ \ \ \ \ \ \ \ \ \ \ \ \ \ \ \ \ \ \ \ \ \ \ \ \ \ $500 \\ 
\hline
 \centering sample color curve & \vspace{4pt}
\includegraphics[width=0.263\textwidth]{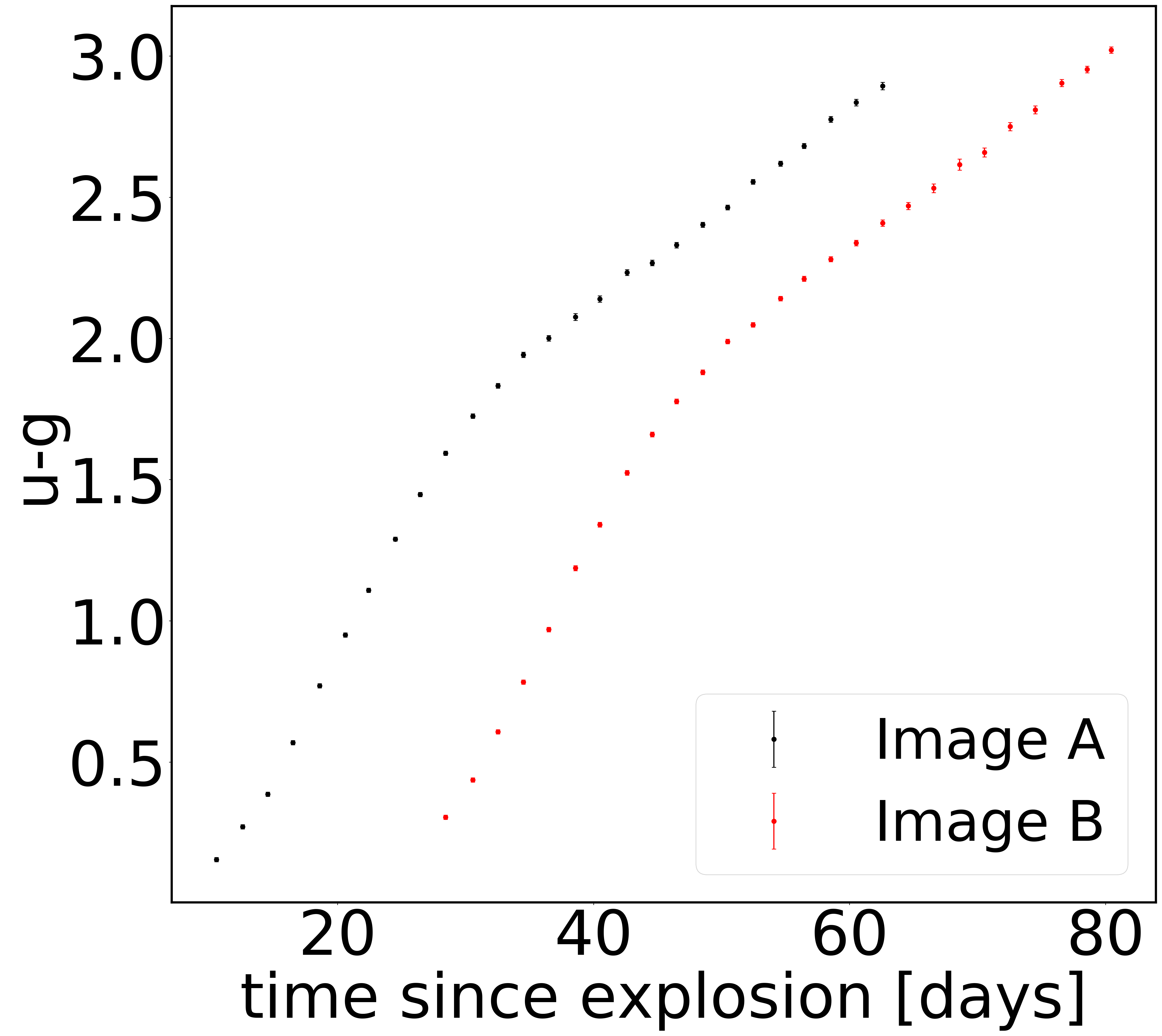}
& \vspace{4pt}
\includegraphics[width=0.25\textwidth]{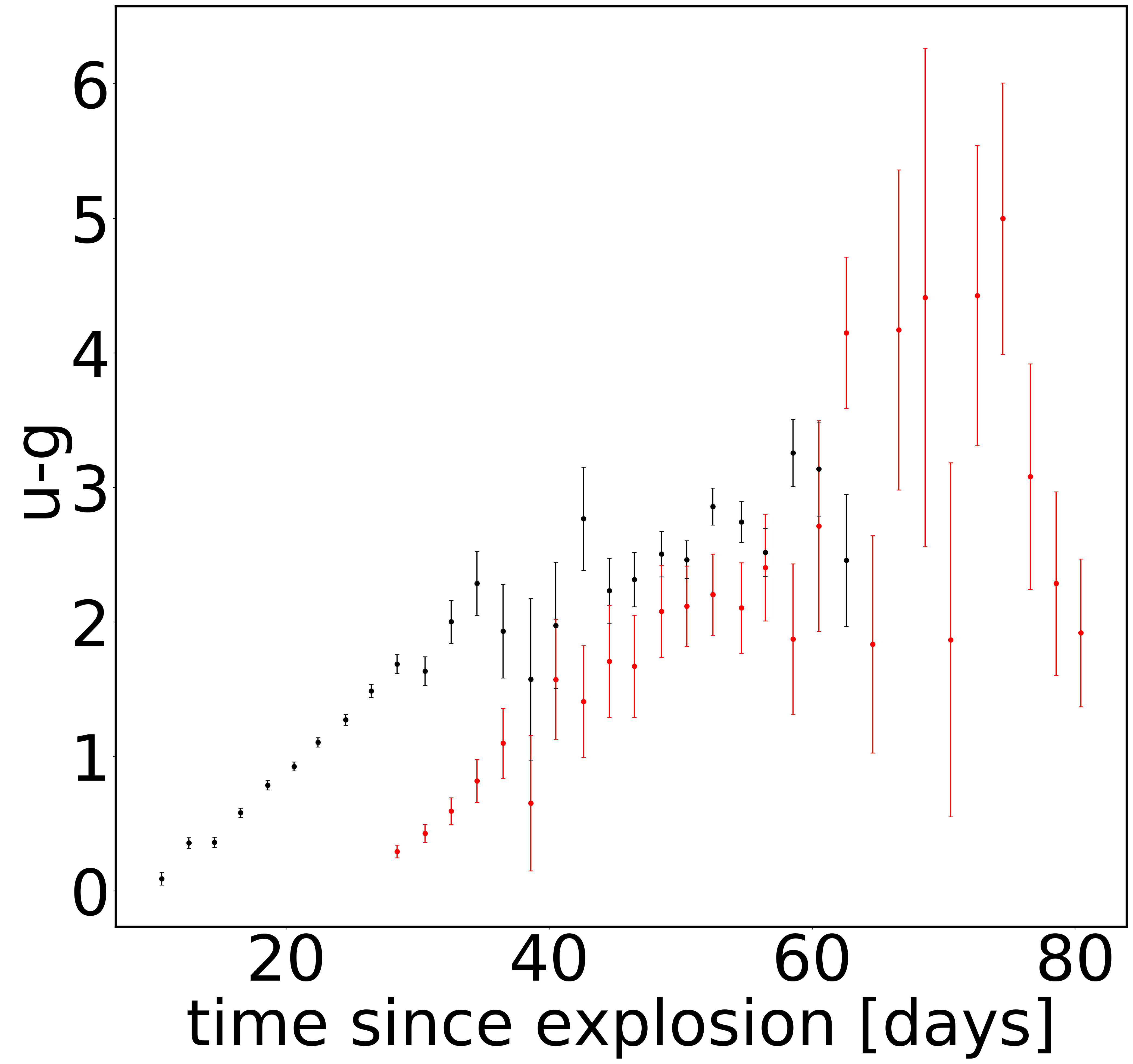}
& \vspace{4pt}
\includegraphics[width=0.25\textwidth]{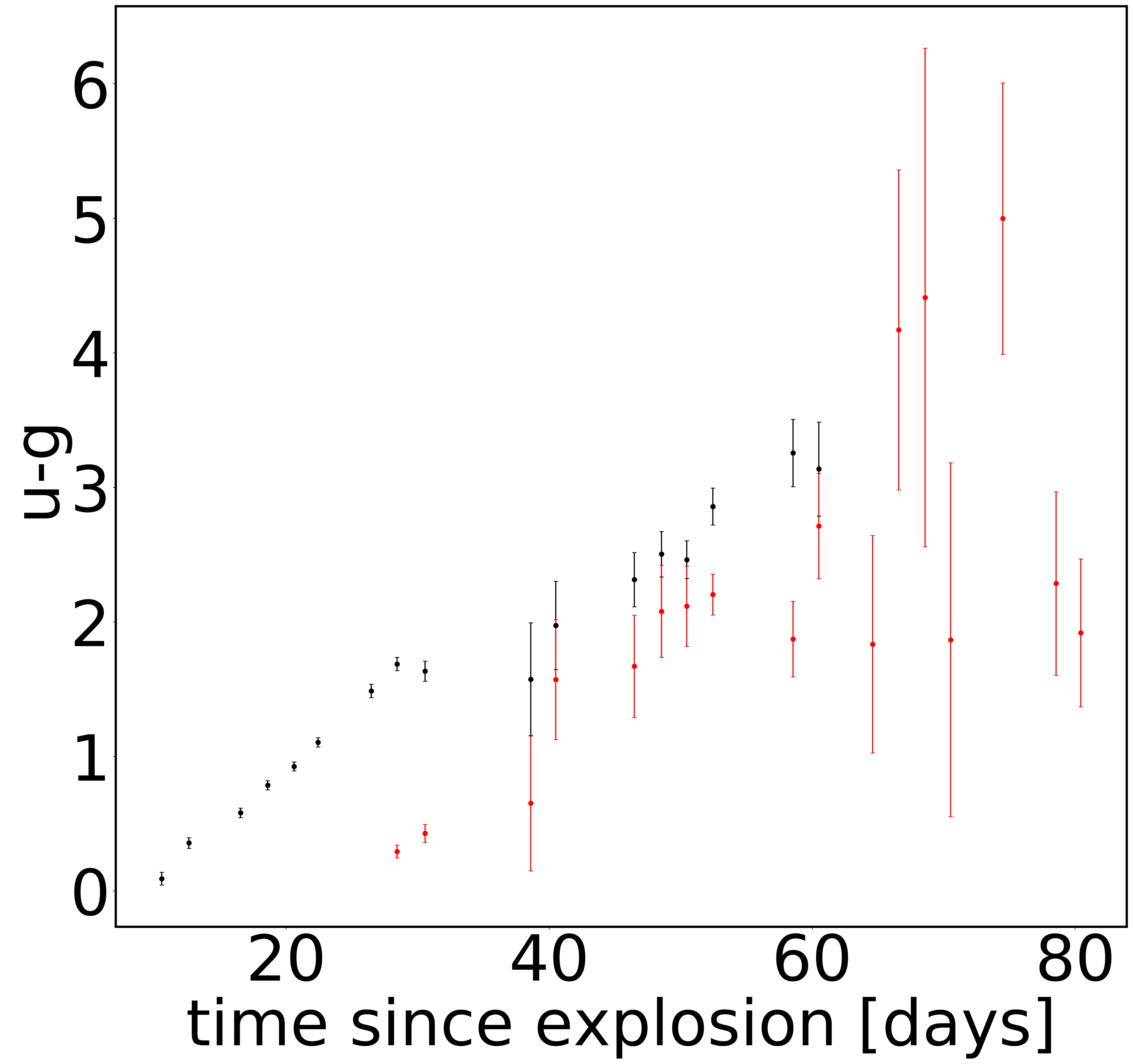}
\\ 
\hline 
\end{tabular} 
\caption{Investigated mock color curves. The top two rows state the color curve scenario and the considered numbers of noise realizations. The third row shows one sample color curve for each scenario including the image A in black and image B in red.}
\label{table_cases}
\end{table*}

We use case 1 just for method-verification purposes, as its high S/N is unrealistic.
To mock up case 2 and case 3, we assume a typical peak absolute brightness at $-$17 mag in the rest-frame V band \citep{Anderson2014} for SNe II. In rare cases the peak absolute magnitude can reach $-$18 mag. Considering a flat $\mathrm{\Lambda}$CDM cosmology with $H_{0} = 72\, \kmsMpc$ \citep{Bonvin2017}, $\Omega_{\rm m}$ = 0.26 and $\Omega_{\rm \Lambda}$ = 0.74 \citep{Oguri2010} and a typical source redshift of lensed SNe II between $z$ = 0.3 and $z$ = 0.6 \citep{Oguri2010}, we get the apparent magnitude range of 23.9 to 25.7 mag.
This is further magnified by the lens with a typical magnification of $\mu$ $\approx$ 3 for the first image and of $\mu$ $\approx$ 1 for the second image \citep{Oguri2010}. The first image will therefore be magnified to around 22.7 magnitudes for the low redshift end and the second image will not be magnified.
To keep our cases 2 and 3 realistic, but also feasible in terms of observational exposure time we choose the V band light curve peak of the first image to be at 22 mag and for the second image at 23 mag and adjust the $u$, $g$, and $r$ accordingly, which is still realistic for bright SNe II at the lower redshift boundary. This could also be achievable observing with a follow up facility with a main mirror larger than 2 m or deeper $5\sigma$ depth than just one magnitude deeper than LSST $5\sigma$ depth.
To quantify the data loss for scenario 3 we analyzed two years of observations of the quasar RXJ1131-1231 at the 1.2 meter Leonhard Euler Telescope with a two-day cadence from COSMOGRAIL. We found that a loss of $\sim$35\% of observations because of bad weather, downtime of the observatory, or the moon phase is a realistic approximation \citep{Millon2020}.
Reddening due to dust is included as discussed in Sect. \ref{sec:MCMC}.

\section{Time-delay and extinction retrieval procedure}
\label{sec:MCMC}

The light curve model in Eq. \ref{eq:5} will also be used in the MCMC sampling process in this section, using lower orders than the full 6 orders.
For the time-delay and extinction parameter retrieval we use a custom Metropolis-Hastings sampling algorithm \citep{Metropolis1953} based on the approach of \cite{Dunkley2005} to ensure full control on the sampling. In this section we introduce our algorithm.

\subsection{Likelihood function}

The likelihood function assigns a probability $p(\vec{\eta}, \Delta t, A_{V,\mathrm{B}}-A_{V,\mathrm{A}})$ to the sampling parameters $\vec{\eta} = (c_{1},c_{2},c_{3},c_{4},c_{5},c_{6},c_{7})_{\mathrm{color \ curve}},\Delta t, A_{V,\mathrm{B}}-A_{V,\mathrm{A}}$, and for Gaussian uncertainties the likelihood can be expressed as $L = A_{\mathrm{norm}} \exp (- \chi^{2} / 2)$ with $A_{\mathrm{norm}}$ being a normalization factor. During the sampling of our MCMC as stated in Sect. \ref{sec:sampling} we are just interested in minimizing the $\chi^{2}$ function and do not regard $A_{\mathrm{norm}}$.
To define $\chi^{2}$ in the most general form, we define the one-dimensional data vectors $\vec{d}_{\mathrm{A}}$ and $\vec{d}_{\mathrm{B}}$ containing the data points of the considered color curves of image A and B. We further consider the model data vectors $\vec{d}_{\mathrm{A}}^{\ \mathrm{model}}$ and $\vec{d}_{\mathrm{B}}^{\ \mathrm{model}}$. The entries of these one-dimensional vectors are defined combining $C(\vec{\eta},t_{j})$ (Eq. \ref{eq:5}) into vector form:
\begin{equation}
d_{\mathrm{A}, j}^{\ \mathrm{model}} = C(\vec{\eta},t_{j})
\end{equation}
and
\begin{equation}
d_{\mathrm{B}, j}^{\ \mathrm{model}} = C(\vec{\eta},t_{j} - \Delta t) + s,
\end{equation}
where $\Delta t$ is the time shift, which we sample together with the parameter vector $\vec{\eta}$ and extinction parameters that are related to $s$.
We write a general shift $s$ in color between image A and B into this definition, which can be expressed by extinction parameters. Considering generic filters $x$ and $y$ of a color curve $x-y$, we define the magnitudes $m$ of each image A and B and each filter, and get:
\begin{equation}\label{eq:shift}
s = (m_{x,\mathrm{B}} - m_{y,\mathrm{B}}) - (m_{x,\mathrm{A}} - m_{y,\mathrm{A}})
\end{equation}
For each image (A or B) and filter ($x$ or $y$), we can write the observed magnitude as:
\begin{equation}\label{eq:magn}
\begin{array}{l}
m_{x,\mathrm{A}} = m_{0,x} - \Delta m_{\mathrm{lens, A}} + A_{x,\mathrm{A}},\\
m_{y,\mathrm{A}} = m_{0,y} - \Delta m_{\mathrm{lens, A}} + A_{y,\mathrm{A}},\\
m_{x,\mathrm{B}} = m_{0,x} - \Delta m_{\mathrm{lens, B}} + A_{x,\mathrm{B}},\\
m_{y,\mathrm{B}} = m_{0,y} - \Delta m_{\mathrm{lens, B}} + A_{y,\mathrm{B}},
\end{array}
\end{equation}
where $m_{0}$ is the intrinsic magnitude, $\Delta m_{\mathrm{lens}}$ is the change from lensing magnification, and $A$ is the extinction coefficient.
Even though both images have different lensing magnifications they have the same $\Delta m_{\mathrm{lens}}$ across filters (i.e. $\Delta m_{\mathrm{lens, A}} = \Delta m_{\mathrm{lens}, x, \mathrm{A}} = \Delta m_{\mathrm{lens}, y, \mathrm{A}}$ and $\Delta m_{\mathrm{lens, B}} = \Delta m_{\mathrm{lens}, x, \mathrm{B}} = \Delta m_{\mathrm{lens}, y, \mathrm{B}}$) as lensing is achromatic, neglecting chromatic microlensing effects, which we expect to be small given that the size of photosphere will not change substantially during the plateau phase \citep{Dessart2011}, which was also confirmed in our earlier microlensing studies in HOLISMOKES V \citep{Bayer2021}. The additional magnification cancels out and the equations from Eq. \ref{eq:magn} can be inserted into Eq. \ref{eq:shift}:
\begin{equation}\label{eq:s}
\begin{array}{l}
s = (m_{0,x} - m_{0,y} - \Delta m_{\mathrm{lens, B}} + \Delta m_{\mathrm{lens, B}} + A_{x,\mathrm{B}} - A_{y,\mathrm{B}})\\
\noindent\hspace*{6mm}%
- (m_{0,x} - m_{0,y} - \Delta m_{\mathrm{lens, A}} + \Delta m_{\mathrm{lens, A}} + A_{x,\mathrm{A}} - A_{y,\mathrm{A}})\\ 
\noindent\hspace*{2.6mm}%
= A_{x,\mathrm{B}} - A_{y,\mathrm{B}} - A_{x,\mathrm{A}} + A_{y,\mathrm{A}}\\[3mm]
\noindent\hspace*{2.6mm}%
= [(a(\lambda_{x}) + \dfrac{b(\lambda_{x})}{R_{V,\mathrm{B}}}) - (a(\lambda_{y}) + \dfrac{b(\lambda_{y})}{R_{V,\mathrm{B}}})] \cdot  A_{V,\mathrm{B}} \\[3mm]
\noindent\hspace*{6mm}%
- [(a(\lambda_{x}) + \dfrac{b(\lambda_{x})}{R_{V,\mathrm{A}}}) - (a(\lambda_{y}) + \dfrac{b(\lambda_{y})}{R_{V,\mathrm{A}}})] \cdot A_{V,\mathrm{A}}\\[3mm]
\noindent\hspace*{2.6mm}%
= [(a(\lambda_{x}) + \dfrac{b(\lambda_{x})}{R_{V}}) - (a(\lambda_{y}) + \dfrac{b(\lambda_{y})}{R_{V}})] \cdot (A_{V,\mathrm{B}}-A_{V,\mathrm{A}})
\end{array}
\end{equation}
For the last two steps we inserted the Cardelli extinction laws from Eq. \ref{eq: Cardelli} and fix $R_{V,\mathrm{A}} = R_{V,\mathrm{B}} = R_{V} = 3.1$ which is the standard value for the diffuse interstellar medium \citep{Cardelli1989}.
The effective wavelengths used to calculate the parameters $a(\lambda)$ and $b(\lambda)$ for the CSP filters are: $\lambda_{u} = 3639.3 \ \mathrm{\AA}$, $\lambda_{g} = 4765.1 \ \mathrm{\AA}$, and $\lambda_{r} = 6223.3 \ \mathrm{\AA}$ \citep{CSP_filters}. 

The final definition of the model data vector for image B is expressed as follows, where $A_{V,\mathrm{B}}-A_{V,\mathrm{A}}$ is the differential extinction parameter:
\begin{equation}
\begin{array}{l}
d_{\mathrm{B},i}^{\ \mathrm{model}} = \vec{C}(\vec{\eta},t_{i} - \Delta t) + [(a(\lambda_{x}) + \dfrac{b(\lambda_{x})}{R_{V}})\\
\noindent\hspace*{13mm}%
- (a(\lambda_{y}) + \dfrac{b(\lambda_{y})}{R_{V}})] \cdot (A_{V,\mathrm{B}}-A_{V,\mathrm{A}}).
\end{array}
\end{equation}

The $\chi^{2}$ is now defined in the most general form as:
\begin{equation}
\begin{array}{l}
\chi^{2} = \chi^{2}_{\mathrm{A}} + \chi^{2}_{\mathrm{B}} \\[1mm]
\noindent\hspace*{4.6mm}%
= (\vec{d}_{\mathrm{A}}-\vec{d}_{\mathrm{A}}^{\ \mathrm{model}})^{T}C_{\mathrm{D},\mathrm{A}}^{-1}(\vec{d}_{\mathrm{A}}-\vec{d}_{\mathrm{A}}^{\ \mathrm{model}}) \\[1mm]
\noindent\hspace*{8mm}%
+ (\vec{d}_{\mathrm{B}}-\vec{d}_{\mathrm{B}}^{\ \mathrm{model}})^{T}C_{\mathrm{D},\mathrm{B}}^{-1}(\vec{d}_{\mathrm{B}}-\vec{d}_{\mathrm{B}}^{\ \mathrm{model}})
\end{array}
\end{equation}
where $C_{\mathrm{D}}$ is the covariance matrix.
For the simple case of only one color curve the covariance matrix has only non zero entries on the main diagonal where the uncertainties of each color curve data point are calculated by adding the uncertainties of the two respective light-curve data points in quadrature. This also holds for using multiple color curves containing only different filters, e.g. for the two color curves $u-g$ and $r-i$, as it is a good approximation to assume that these filters do not correlate.
For the case of using color curves containing the same filter, e.g. the two color curves $u-g$ and $u-r$, we also have entries on the diagonals within the off-diagonal blocks of the matrix, which are not zero.
The inverse of the covariance matrix is calculated using the Cholesky decomposition of the numpy.linalg python module \citep{Harris2020}.

\subsection{Sampling technique and prior}
\label{sec:sampling}

We start to sample the posterior probability distribution with initial parameters $\vec{\eta}_{0}, \Delta t_{0}$ and $(A_{V,\mathrm{B}}-A_{V,\mathrm{A}})_{0}$ selected with a $\chi^{2}$ non-linear fit of the mock color curves based on a Levenberg-Marquart algorithm as implemented in the {\texttt optimize.curve\_fit} function of the python module {\texttt scipy} \citep{2020SciPy-NMeth}. 
The next set of parameters $\vec{\eta}_{i}$ of step $i$ is selected by the walker setting the step size. For the initial chain we define the trial probability distribution for taking the next step as a multivariate Gaussian with the identity matrix as the covariance matrix. For the subsequent chain we use the covariance matrix calculated from the previous chain, and scale it by a constant factor to obtain the trial probability distribution with the desired acceptance rate. This ensures that we have an ideal step size of the walker and an acceptance rate around $\sim$25\% after several chains. 
We apply a broad uniform prior of [$-$100,100] for the time delay. All other parameters have improper uniform priors.
To sample the parameters $\left\lbrace \vec{\eta}, \Delta t, A_{V,\mathrm{B}}-A_{V,\mathrm{A}} \right\rbrace $, we follow the Metropolis-Hastings algorithm \citep{Dunkley2005}. 
We perform a check for convergence as described in Sect. \ref{sec:conv_test}, and if convergence is not achieved, we keep running the chain.

\subsection{Convergence testing}
\label{sec:conv_test}

During the convergence test, the power spectrum of the samples of each parameter is fitted with the parametric function
\begin{equation}
P(k) = P_{0}\frac{(N k^{*} / 2 \pi j)^{\alpha}}{1 + (N k^{*} / 2 \pi j)^{\alpha}}.
\end{equation}
from \cite{Dunkley2005}. It considers the chain length $N$, the wave number $k^{*}$ at the turn-over of the power spectrum, the power $P_{0}$ at wave number $k = 0$, the Fourier modes $j = k (N / (2 \pi))$, and an exponent $\alpha$.
A chain has converged if the following two criteria are met:\\
\noindent\hspace*{8mm}%
1. $j^{*} > 20 $, with $j^{*} = k^{*} (N / (2 \pi))$ as the turn-over Fourier\\
\noindent\hspace*{11mm}
mode. This ensures that the minimum wave number $k_{\mathrm{min}}$\\
\noindent\hspace*{11mm} 
is in the regime of white noise $P(k) \sim k^{0}$ indicating that\\
\noindent\hspace*{11mm} 
the region of high probability has been fully sampled.\\
\noindent\hspace*{8mm}%
2. The convergence ratio $r$ for normalized chains should\\
\noindent\hspace*{11.5mm} 
obey $r = (P_{0} / N)< 0.01$.\\
If one of the criteria is not met, we explore additional 20000 steps before testing again and repeat this process until convergence is reached. The first 20\% of the converged chain are removed as the burn-in phase before convergence testing.

\subsection{Polynomial order selection with the Bayesian Information Criterion}

To have a statistically significant result, we investigate several noise realizations for each mock case. For case 1 of Table \ref{table_cases} we use 100 noise realizations and for the cases 2 and 3 we use 500 noise realizations, as fluctuations increase with lower S/N. 
Each noise realization is fitted with different orders of the orthogonal polynomials as presented in Section \ref{sec:light_curve_model}, ranging from the linear order 1 to the most complex case of order 6. To select which order to include in the final results we apply the Bayesian Information Criterion (BIC) to each noise realization and fitted order. The BIC is a simplified criterion for Bayesian model comparison and defined as:
\begin{equation}
\mathrm{BIC} = \ln(N_{\mathrm{data}})N_{\mathrm{par}} - 2 \ln(\hat{L})
\end{equation}
with the number of data points $N_{\mathrm{data}}$, number of free parameters $N_{\mathrm{par}}$ and maximum likelihood $\hat{L}$ of the noise realization \citep{Schwarz1978}. Choosing the model with the lowest BIC ensures that not only the best fitting model is considered, but also the complexity of each model is taken into account.
For the final combination of all investigated noise realizations, we consider the polynomial order of fitting with the lowest BIC value for each realization.

\section{Time-delay and extinction retrieval results}
\label{sec:results}

In this section, we present the precision and accuracy of our time shift and differential extinction parameter determination method with lensed SN IIP color curves. The input values in the rest-frame mock data creation for all cases in this section are $\Delta t$ = 16.3 days and $A_{V,\mathrm{B}}-A_{V,\mathrm{A}}$ = 0.2 mag, which is a realistic value based on the study of the extinction properties of mostly elliptical lensing galaxies \citep{Eliasdottir2006}.
For each of the scenarios 1, 2, and 3 introduced in Sect. \ref{sec:color_curves} and Table \ref{table_cases} we investigate the single color curve $u-g$ as well as the combinations of two color curves $u-g$ and $u-r$. 
In Appendices \ref{App:diff_model}, \ref{lin_case}, \ref{diff_shift}, and \ref{RV} we investigate additional scenarios to test the robustness of our method including other time-shift values.

\begin{figure*}[hbt!]
\centering
\subfigure[$6^{\mathrm{th}}$ order]{\label{}\includegraphics[width=0.3\textwidth]{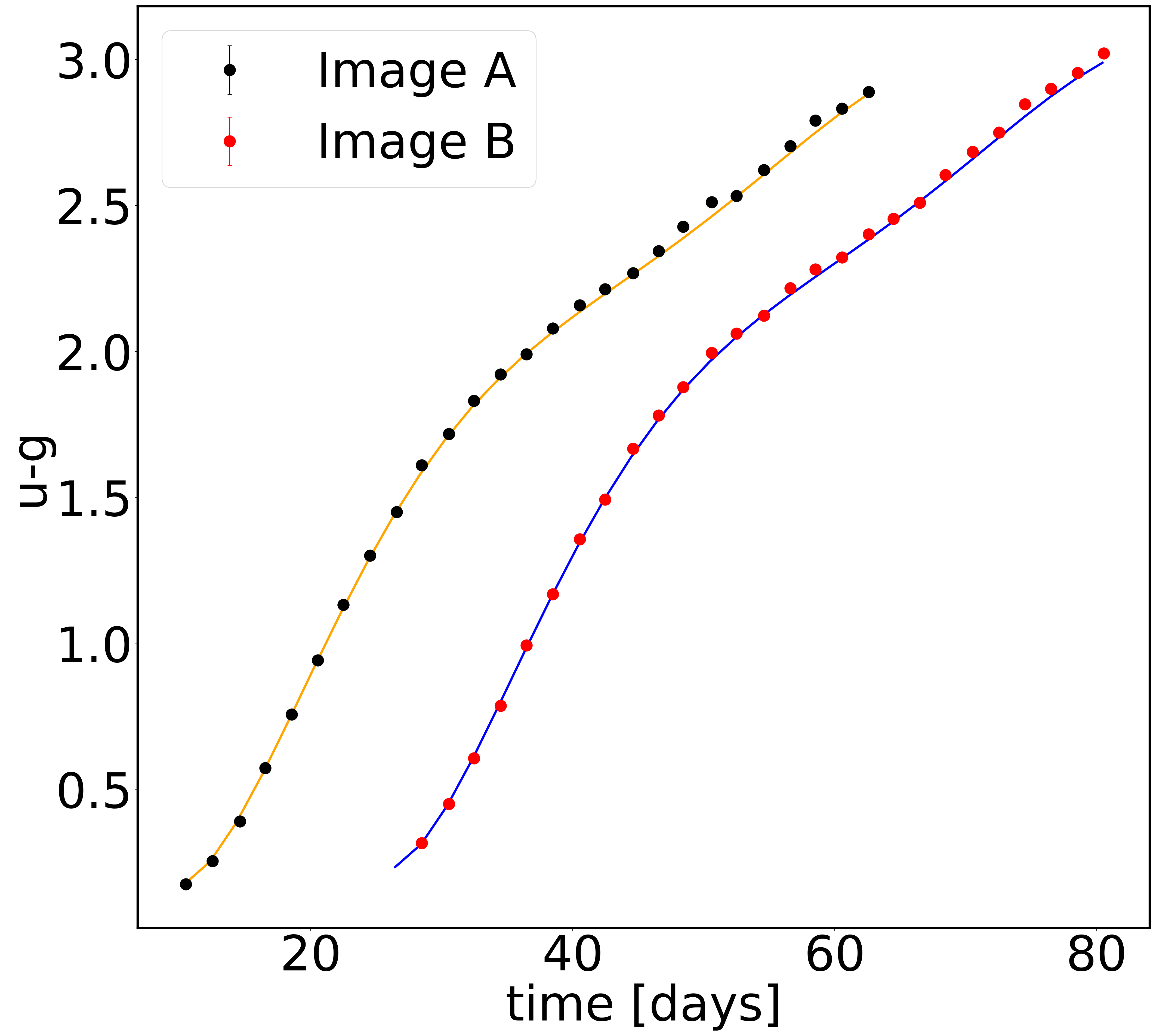}}
\noindent\hspace*{1mm}%
\subfigure[$\mathbf{5^{\mathrm{th}}}$ \textbf{order}]{\label{}\includegraphics[width=0.3\textwidth]{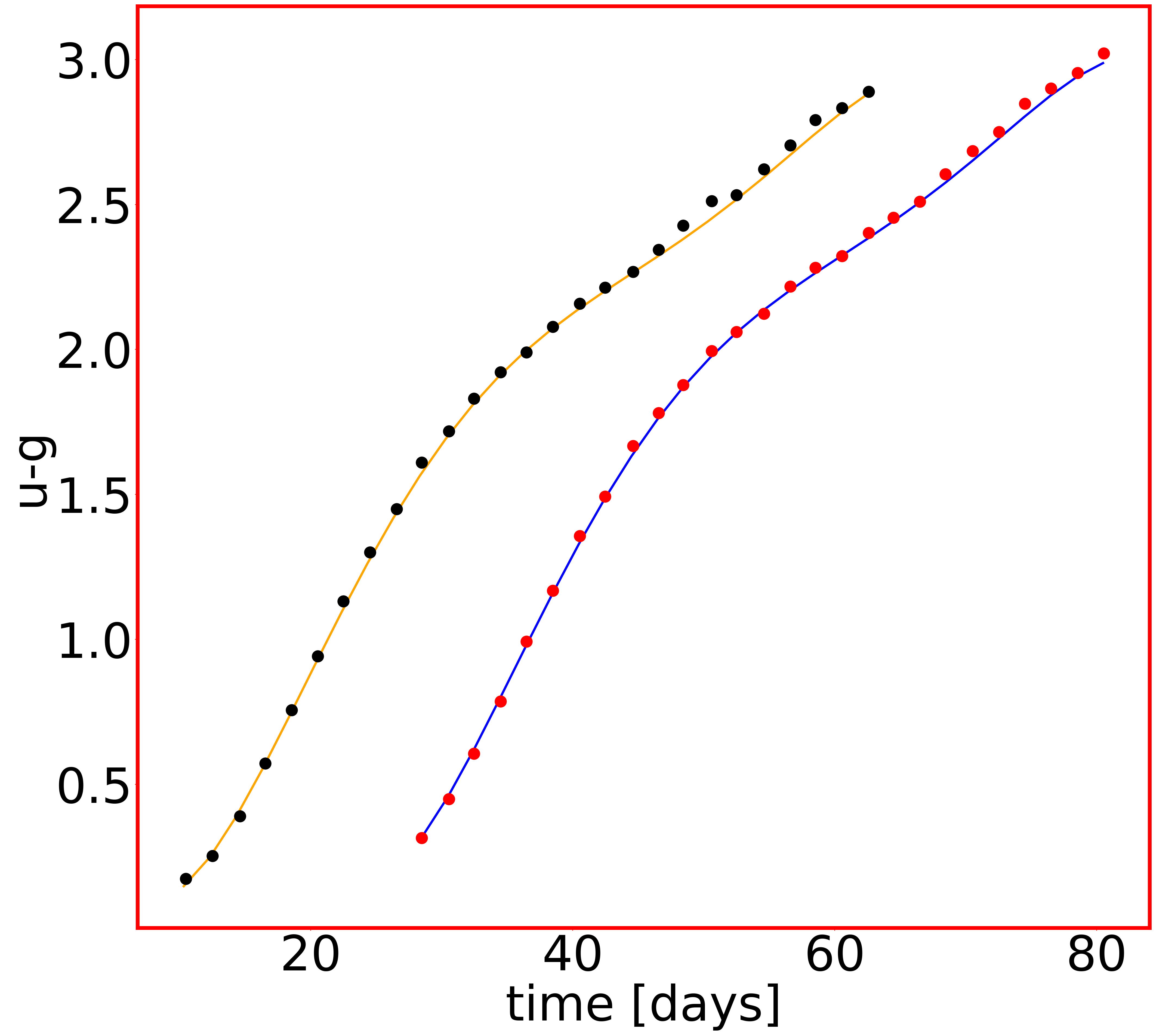}}\\
\subfigure[$4^{\mathrm{th}}$ order]{\label{}\includegraphics[width=0.3\textwidth]{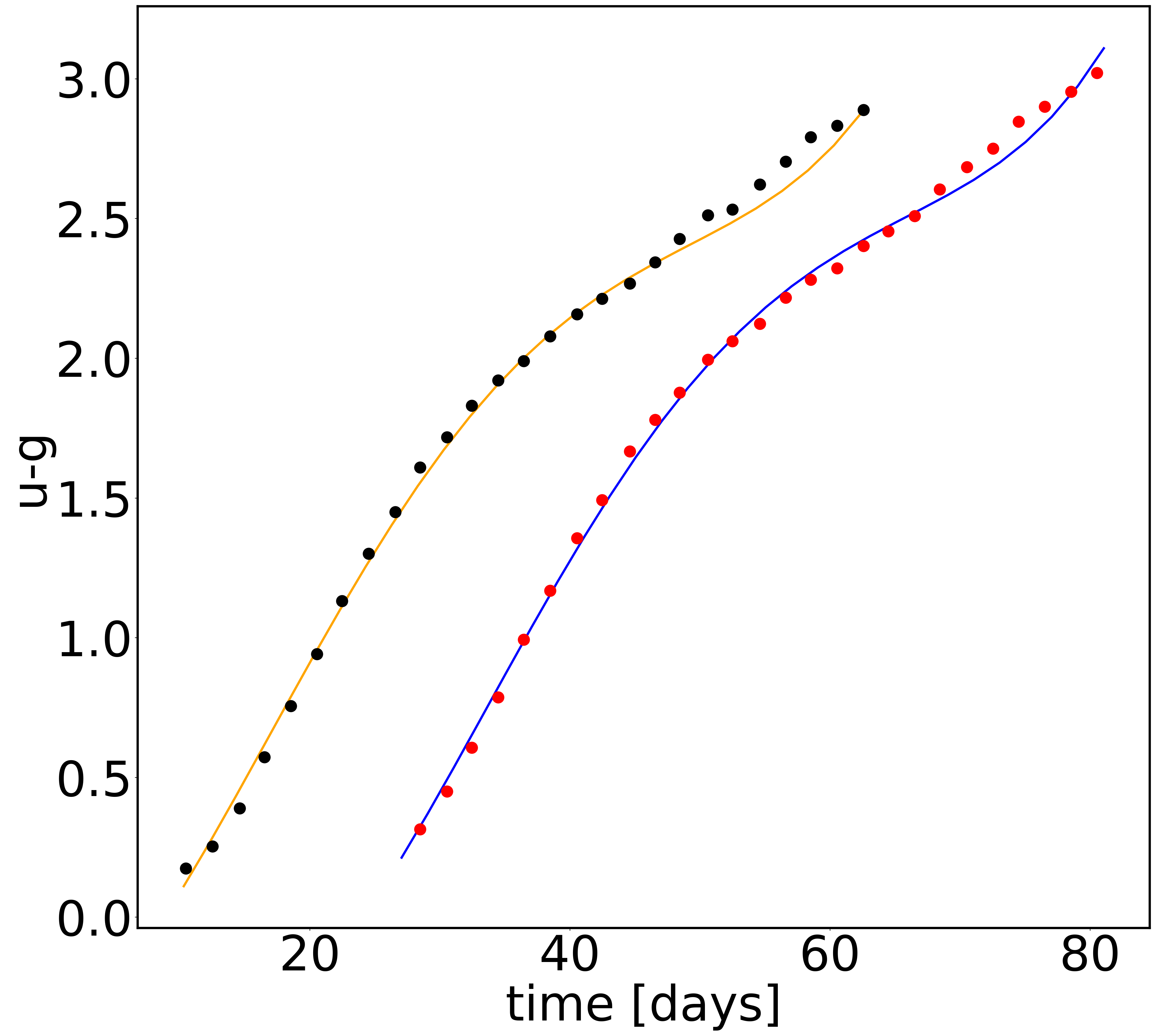}}
\noindent\hspace*{1mm}%
\subfigure[$3^{\mathrm{rd}}$ order]{\label{}\includegraphics[width=0.3\textwidth]{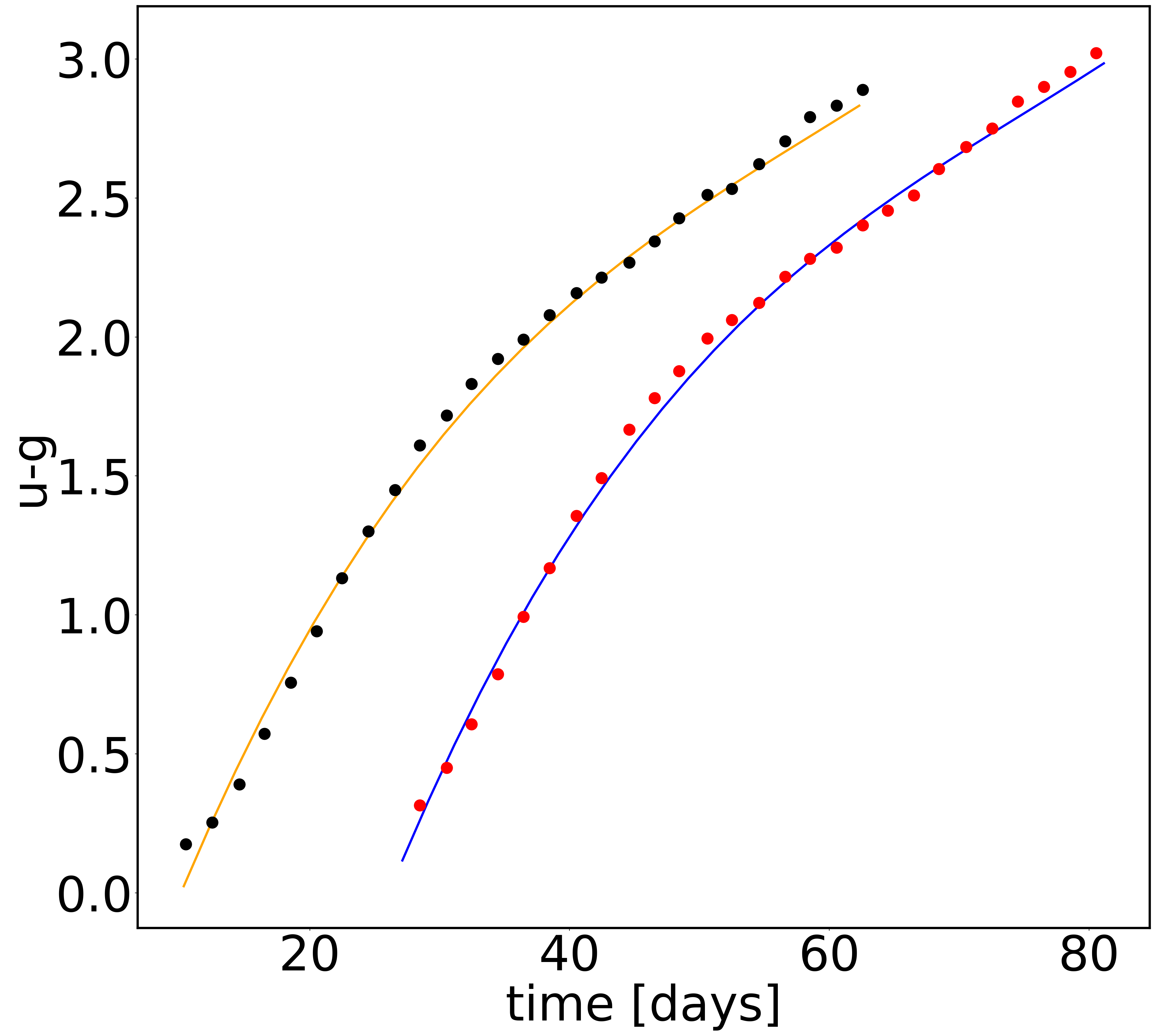}}
\subfigure[$2^{\mathrm{nd}}$ order]{\label{}\includegraphics[width=0.3\textwidth]{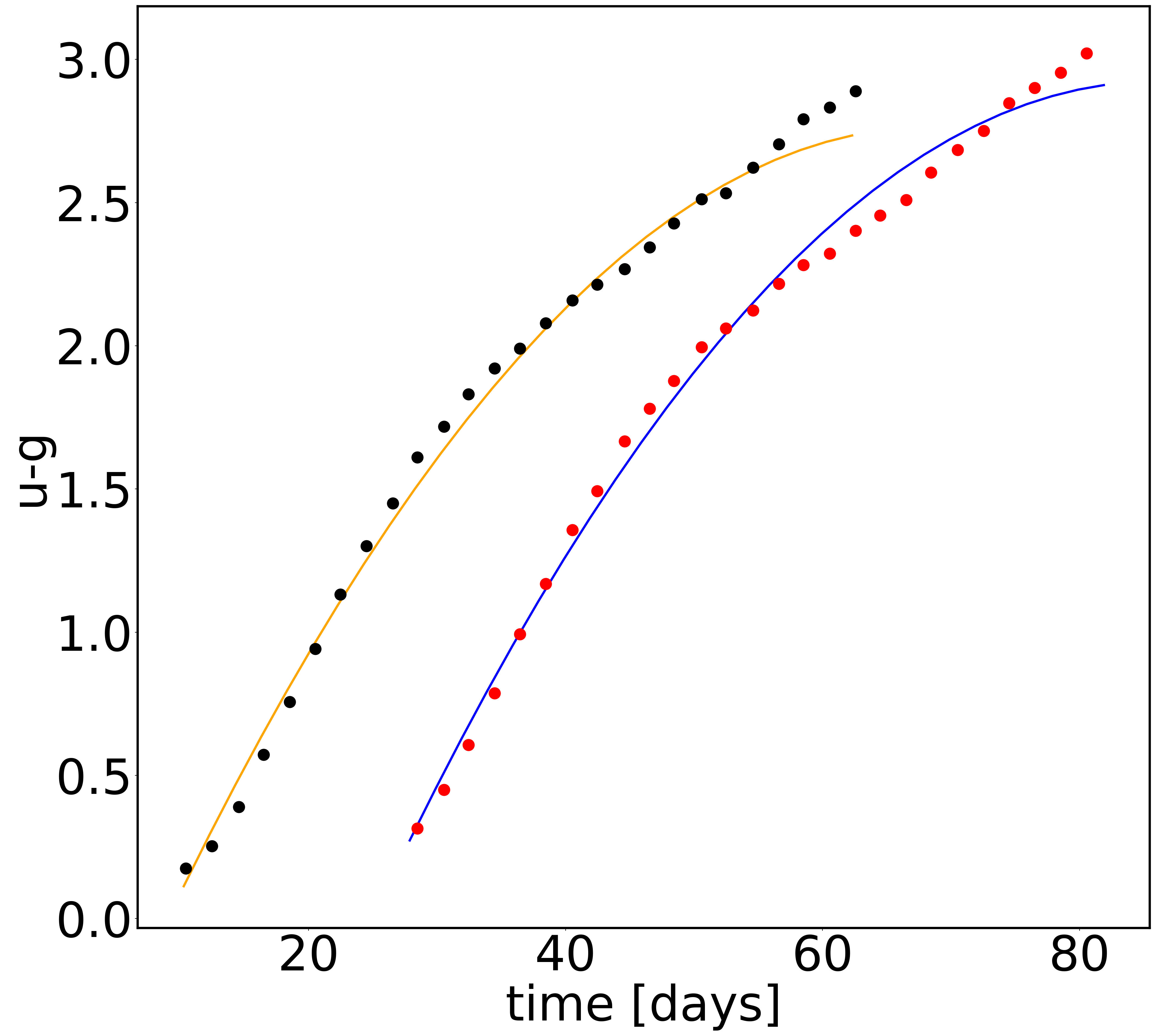}}\\\caption{\label{BIC1}Orthogonal polynomial fits to one noise realization of case 1 to demonstrate the BIC selection. The selected order is highlighted with bold letters and a red frame of the plot.}
\end{figure*}
\begin{figure*}[hbt!]
\centering
\subfigure[$6^{\mathrm{th}}$ order]{\label{}\includegraphics[width=0.3\textwidth]{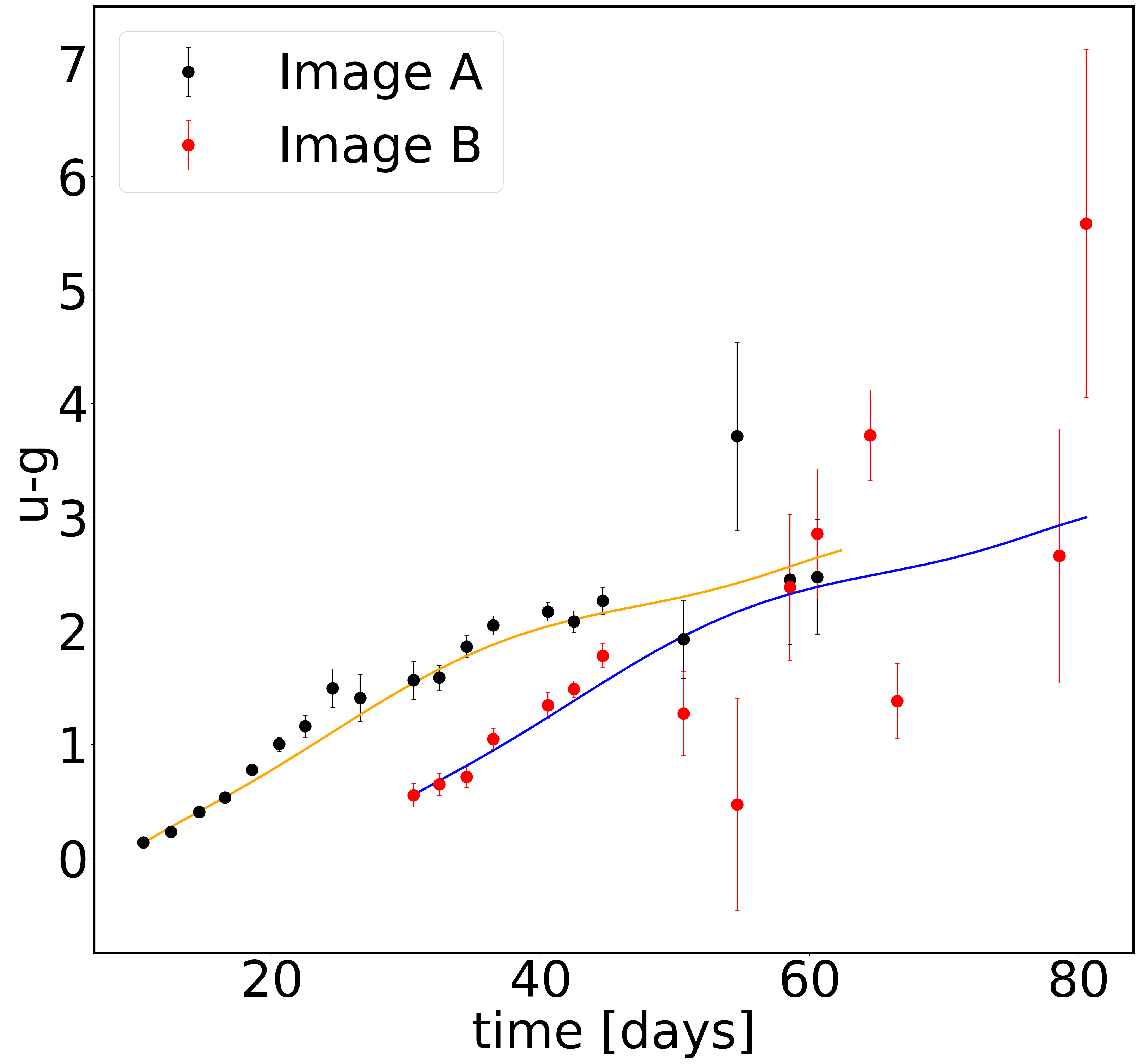}}
\noindent\hspace*{1mm}%
\subfigure[$5^{\mathrm{th}}$ order]{\label{}\includegraphics[width=0.3\textwidth]{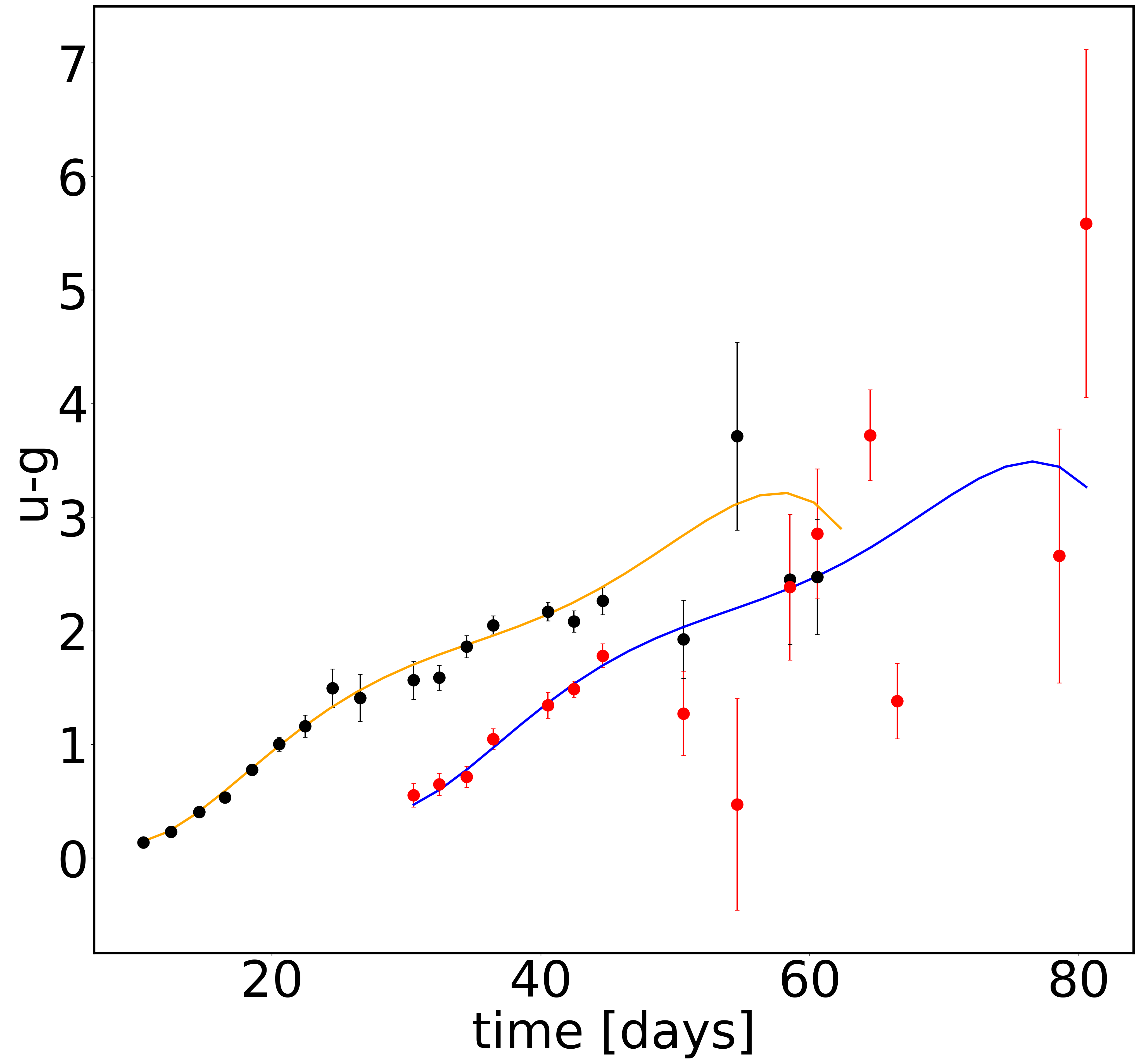}}\\
\subfigure[$\mathbf{4^{\mathrm{th}}}$ \textbf{order}]{\label{}\includegraphics[width=0.3\textwidth]{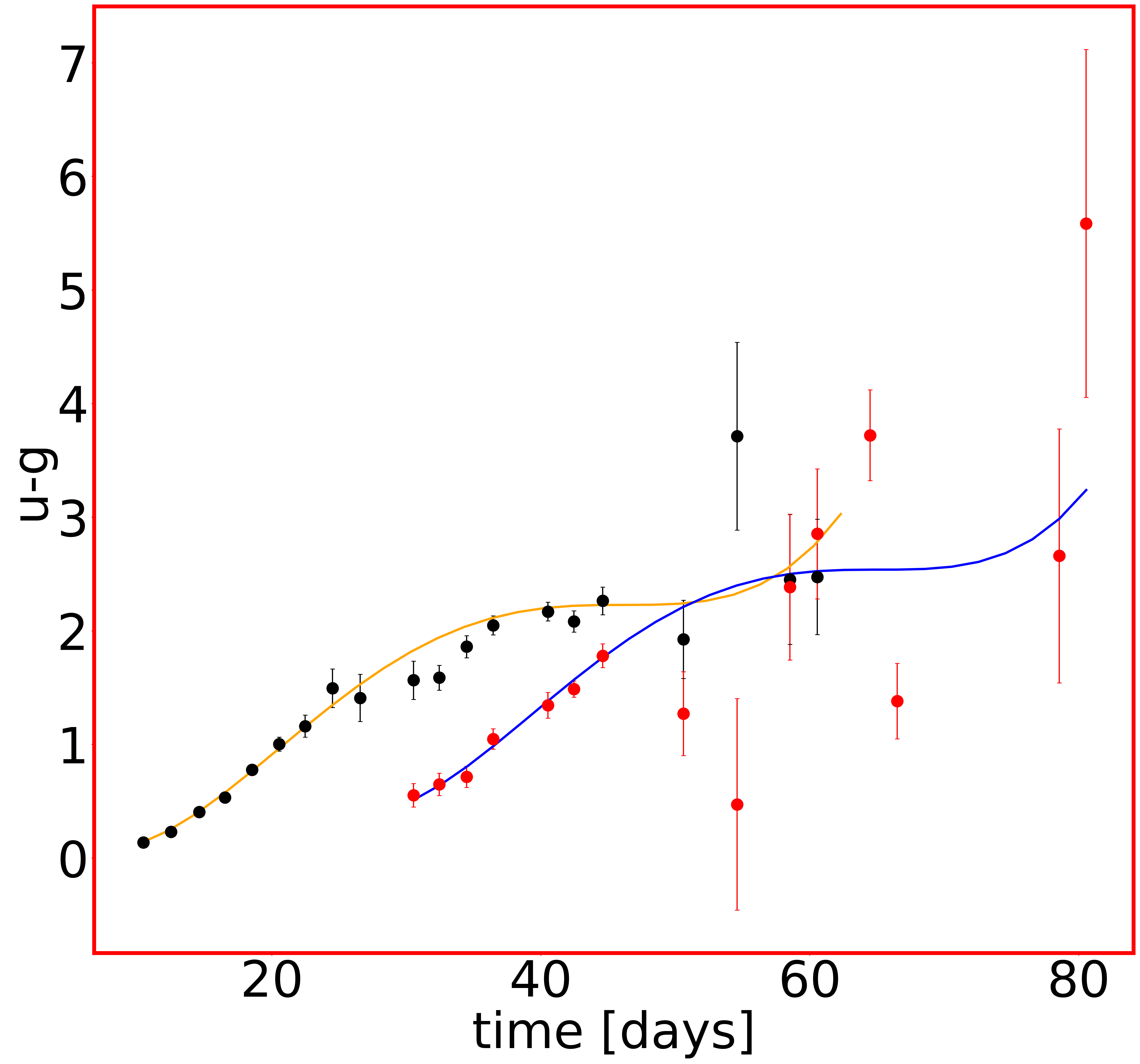}}
\noindent\hspace*{1mm}%
\subfigure[$3^{\mathrm{rd}}$ order]{\label{}\includegraphics[width=0.3\textwidth]{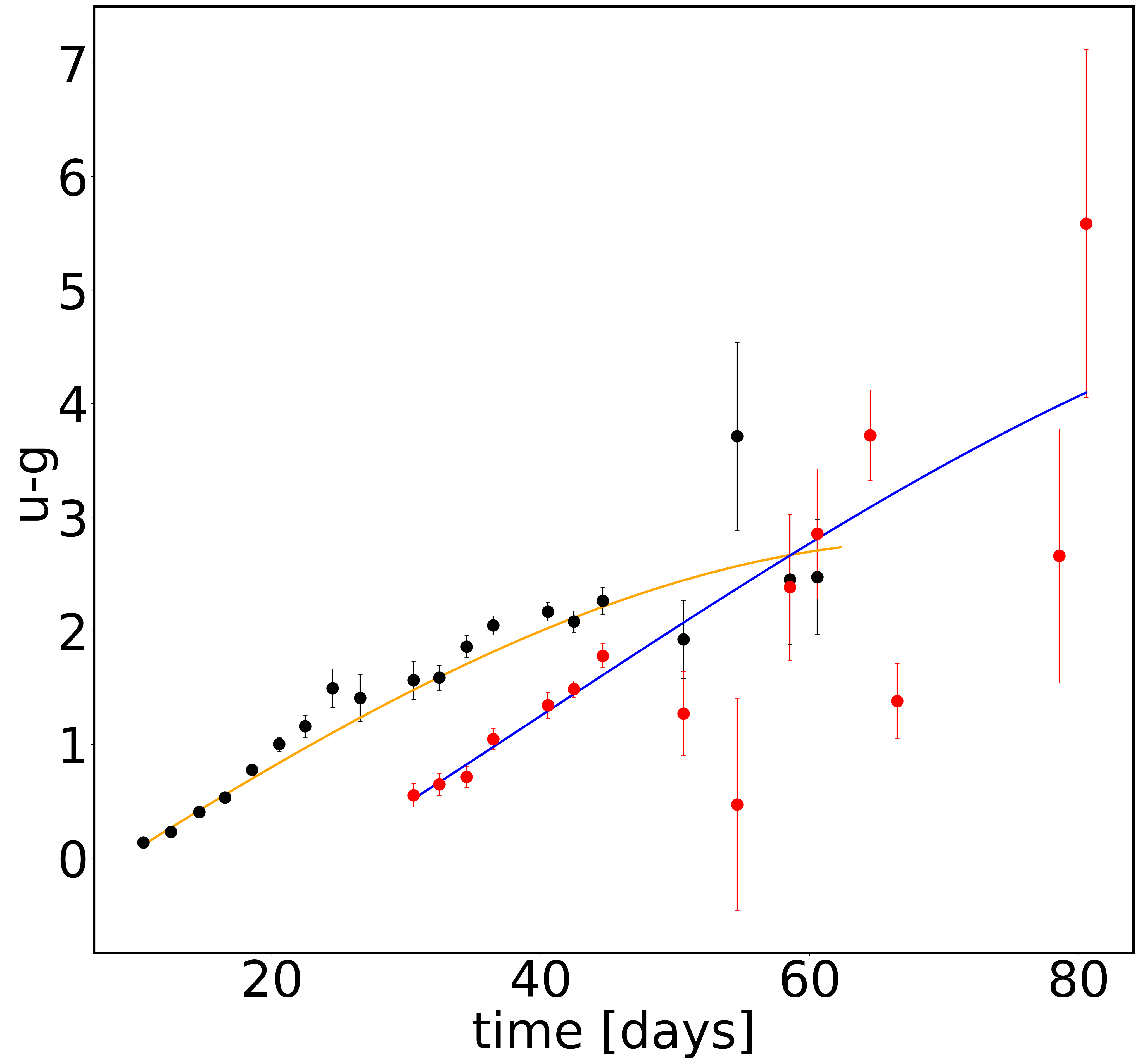}}
\subfigure[$2^{\mathrm{nd}}$ order]{\label{}\includegraphics[width=0.3\textwidth]{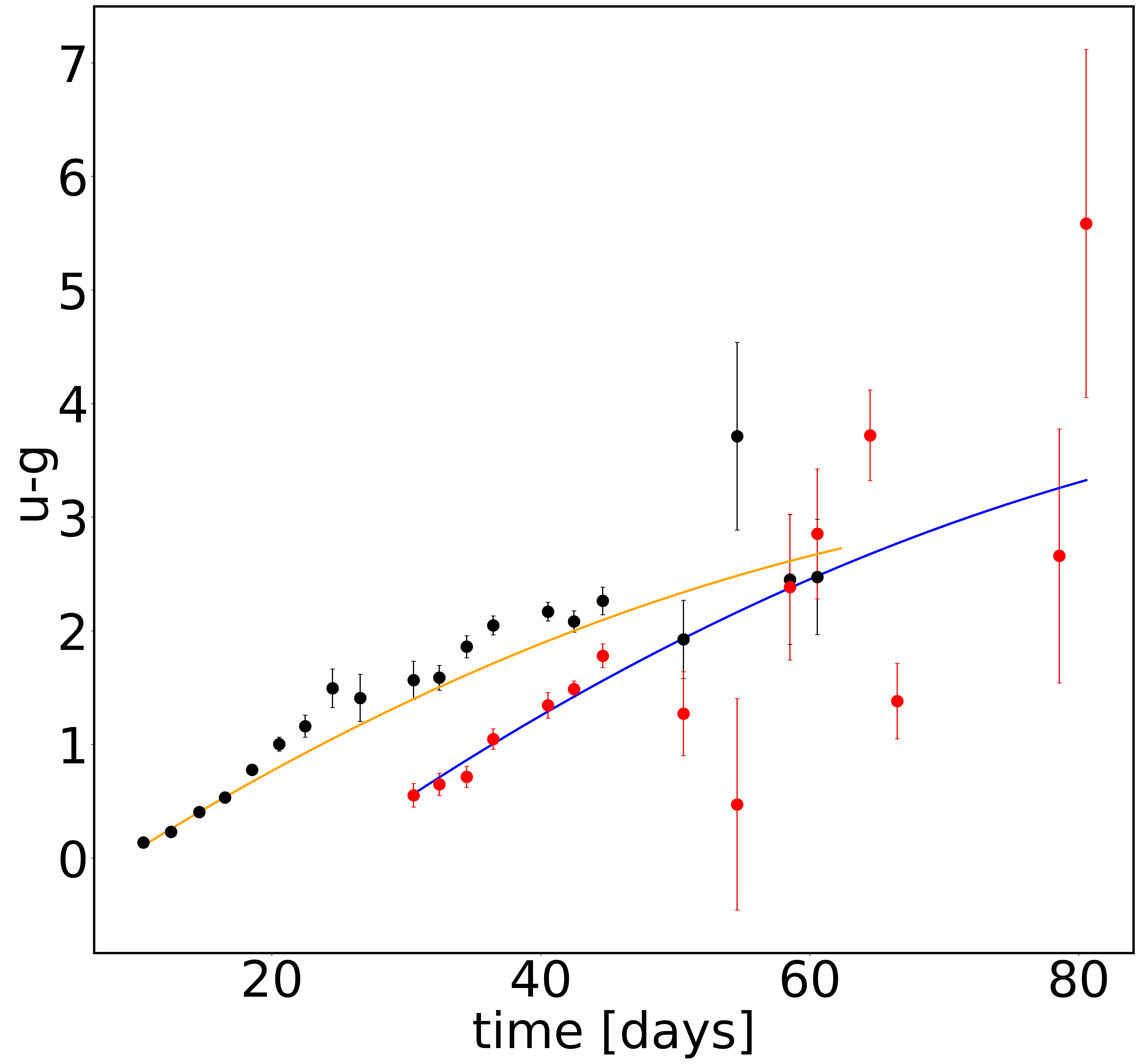}}\\\caption{\label{BIC3}Orthogonal polynomial fits to one noise realization of case 3 to demonstrate the BIC selection. The selected order is highlighted with bold letters and a red frame of the plot. For the $3^{\mathrm{rd}}$ order the retrieved time delay and differential dust extinction deviate strongly from the input which is visible from the different shapes of the fits in the range of the data points of image A and image B.}
\end{figure*}

\begin{table*}[t]
\centering
\begin{tabular}{|m{0.8cm}||m{2.6cm}|m{3.9cm}|}
\hline \centering order & \centering Case 1: high S/N & Case 3: low S/N, 35\% data loss \\
\hline\hline
\centering 6 & 97.4 & 70.0 \\
\centering 5 & \textbf{88.5} & 50.6 \\
\centering 4 & 1090.8 & \textbf{45.2} \\
\centering 3 & 1498.6 & 56.5 \\
\centering 2 & 1859.7 & 63.4 \\
\hline 
\end{tabular} 
\caption{BIC values for the noise realizations shown in Fig. \ref{BIC1} and Fig. \ref{BIC3}. The selected BIC values are highlighted in bold numbers.}
\label{BIC_table}
\end{table*}

Before presenting the results we demonstrate the selection of the orders via the BIC for one case 1 and one case 3 noise realization. The high S/N case 1 mock data example with fits of orders 6 down to 2 are shown in Fig. \ref{BIC1}, where we highlight the selected order with a red frame. Further we list the corresponding BIC values in Table \ref{BIC_table}. It is visible that the 5$^{\mathrm{th}}$ order fits as well as the 6$^{\mathrm{th}}$ order, and is favored in the BIC calculation. The lower orders visibly fit worse.
For case 3 we show the same orders fit to the mock data in Fig. \ref{BIC3} as well as the BIC values in Table \ref{BIC_table}. For this case it is not clearly visible which order to favor, but considering the calculated BIC value the 4$^{\mathrm{th}}$ order is selected for this specific noise realization.
For each case this procedure is applied to all noise realizations resulting in a combination of fits with different orders for the full data set.

\begin{figure*}[hbt!]
\centering
\subfigure[Case 1 with one color curve $u-g$.]{\label{}\includegraphics[width=0.47\textwidth]{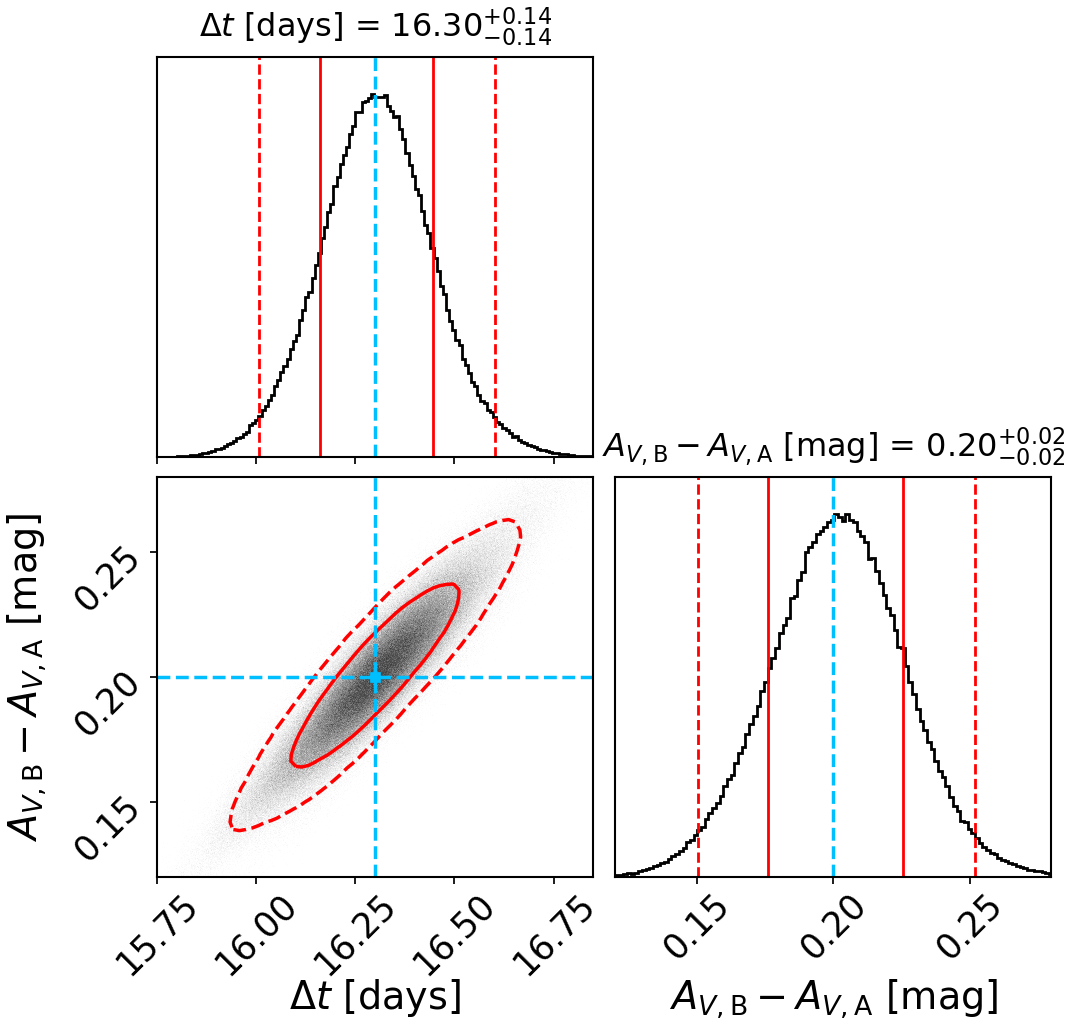}}
\noindent\hspace*{1mm}%
\subfigure[Orders of fitting selected for case 1 with one color curve $u-g$.]{\label{bar_1_1}\includegraphics[width=0.47\textwidth]{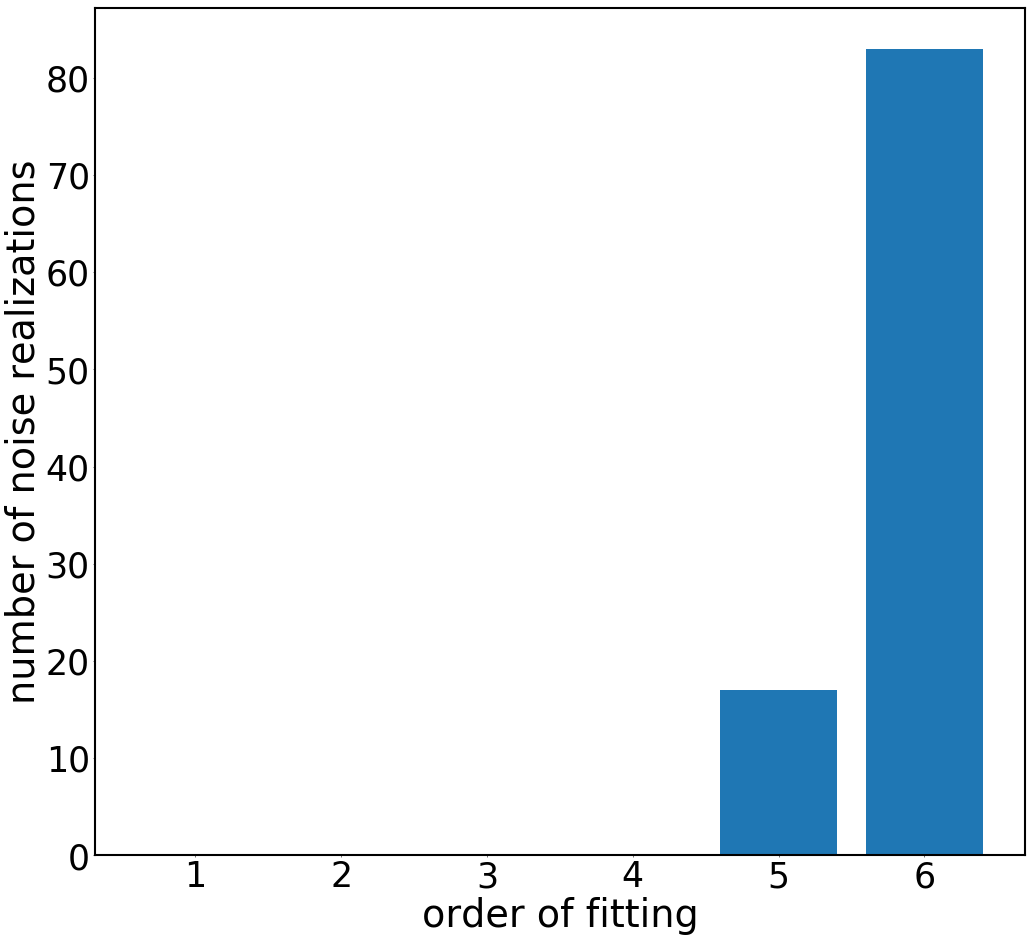}}\\
\subfigure[Case 1 with two color curves $u-g$ and $u-r$.]{\label{}\includegraphics[width=0.47\textwidth]{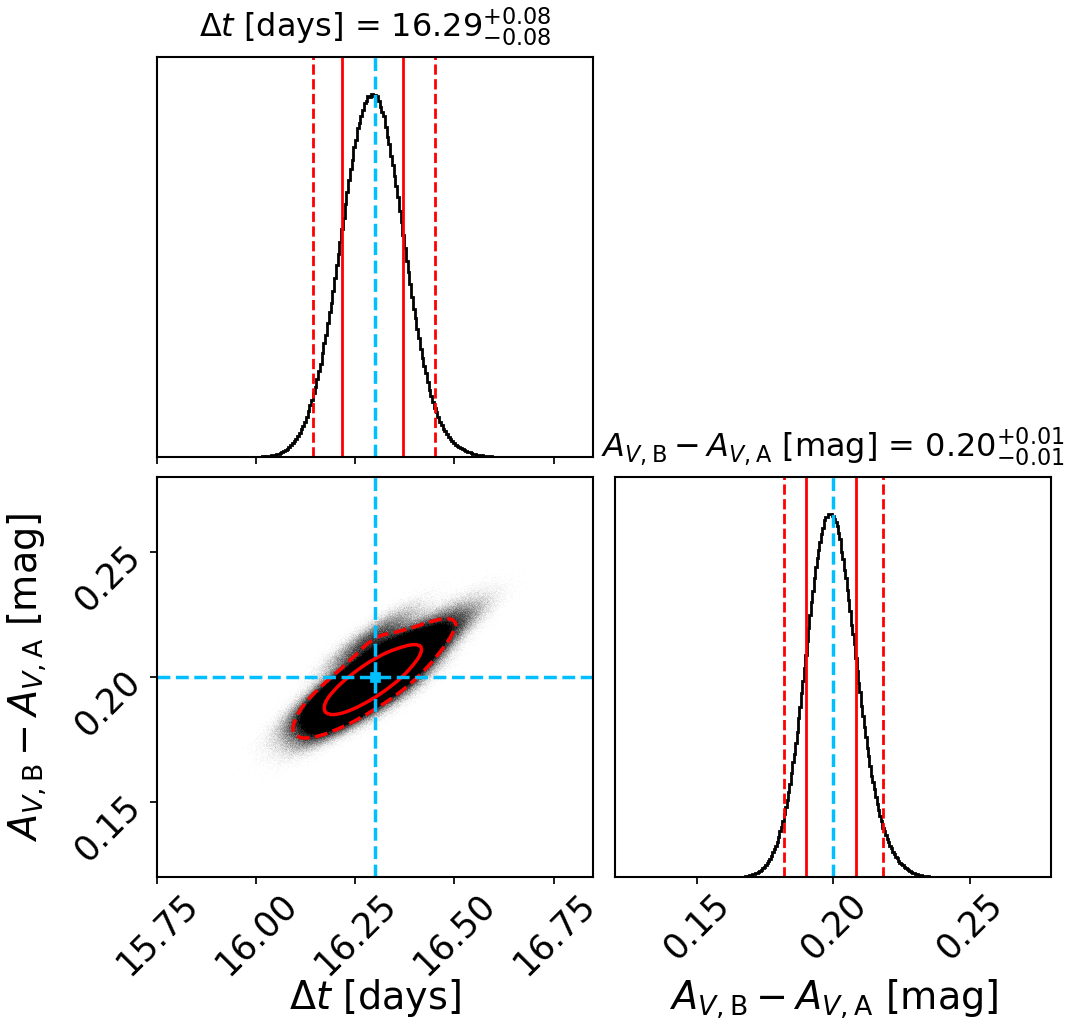}}
\noindent\hspace*{1mm}%
\subfigure[Orders of fitting selected for case 1 with two color curves $u-g$ and $u-r$. We impose the same order for both color curves.]{\label{bar_1_2}\includegraphics[width=0.47\textwidth]{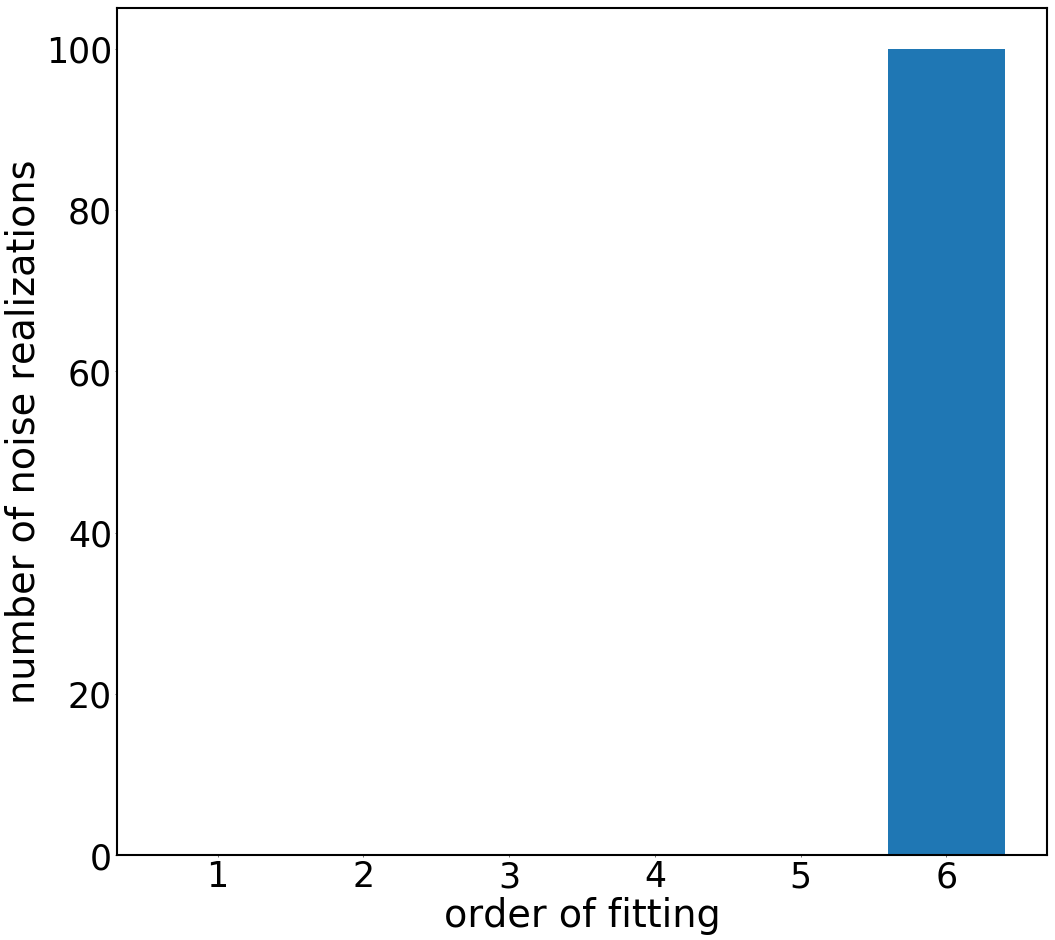}}\\\caption{\label{SN2005J_corner1}The left panels show corner plots of the time shift and differential dust extinction retrieved from 100 noise realizations for case 1. We plot the contours for the 1$\sigma$ (68\%, solid red line) and 2$\sigma$ (95\%, dashed red line) credible intervals for the differential dust extinction $A_{V,\mathrm{B}}-A_{V,\mathrm{A}}$ and the time delay $\Delta t$. The input values in light blue are $\Delta t$ = 16.3 days and $A_{V,\mathrm{B}}-A_{V,\mathrm{A}}$ = 0.2 mag.
On the right we show histograms of the selected orthogonal polynomial orders of fitting using the BIC for the corresponding cases on the left.}
\end{figure*}

\begin{figure*}[hbt!]
\centering
\subfigure[Case 2 with one color curve $u-g$.]{\label{}\includegraphics[width=0.49\textwidth]{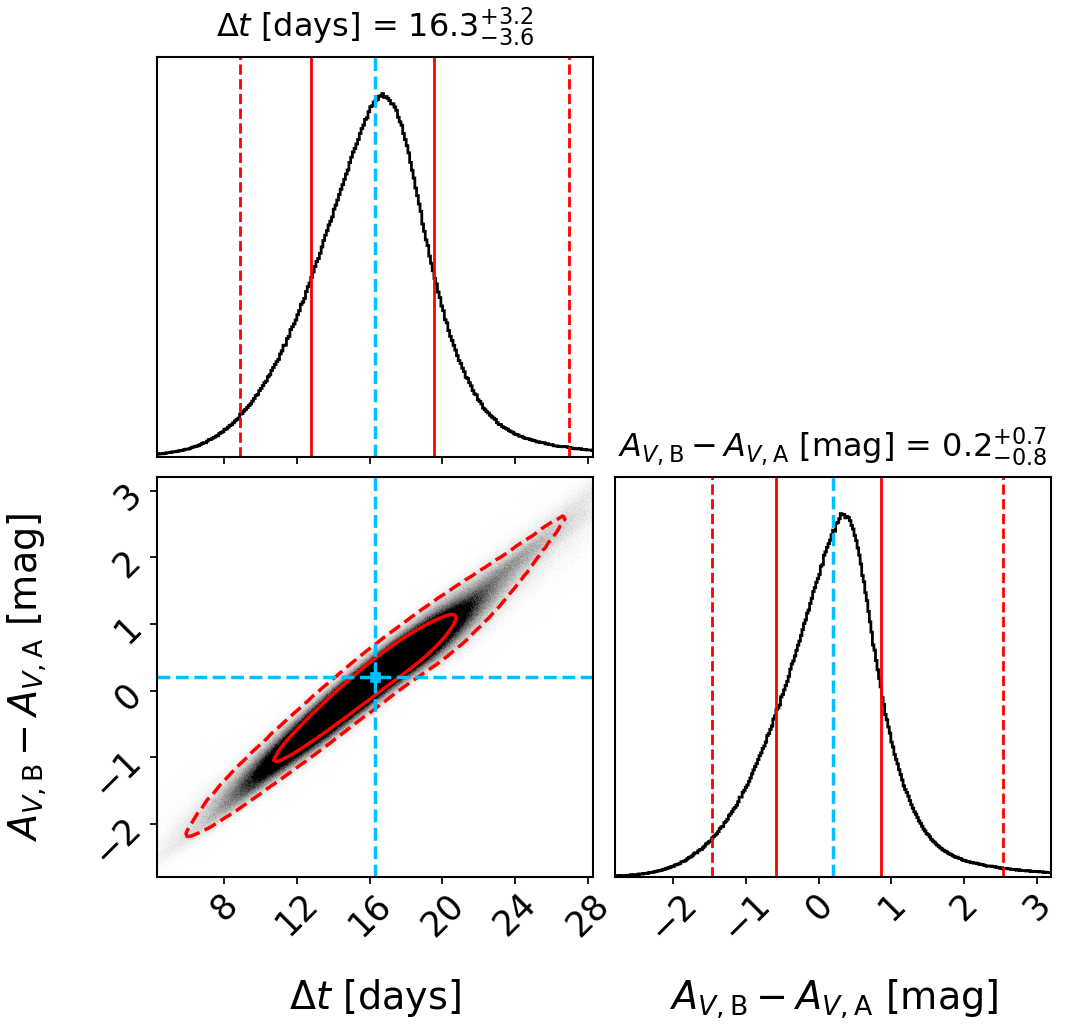}}\noindent\hspace*{1mm}%
\subfigure[Orders of fitting selected for case 2 with one color curve $u-g$.]{\label{}\includegraphics[width=0.49\textwidth]{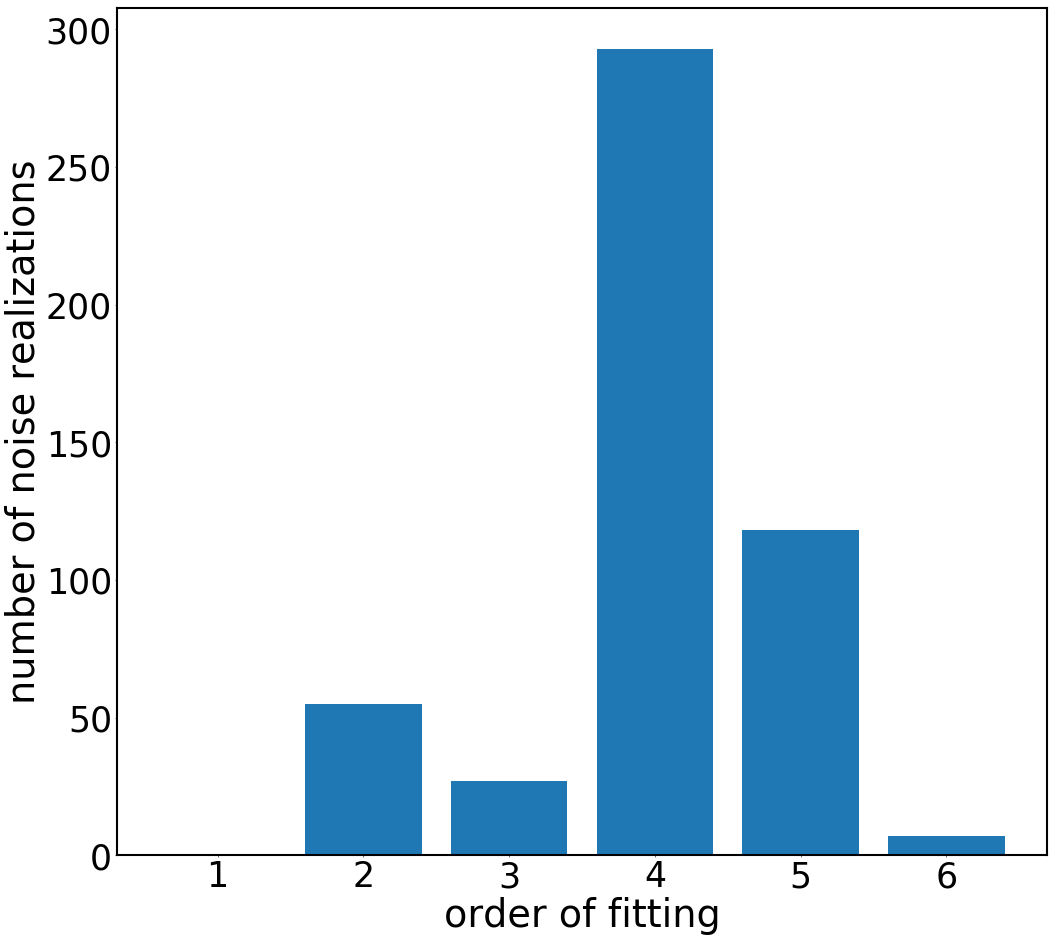}}\\
\subfigure[Case 2 with two color curves $u-g$ and $u-r$.]{\label{}\includegraphics[width=0.49\textwidth]{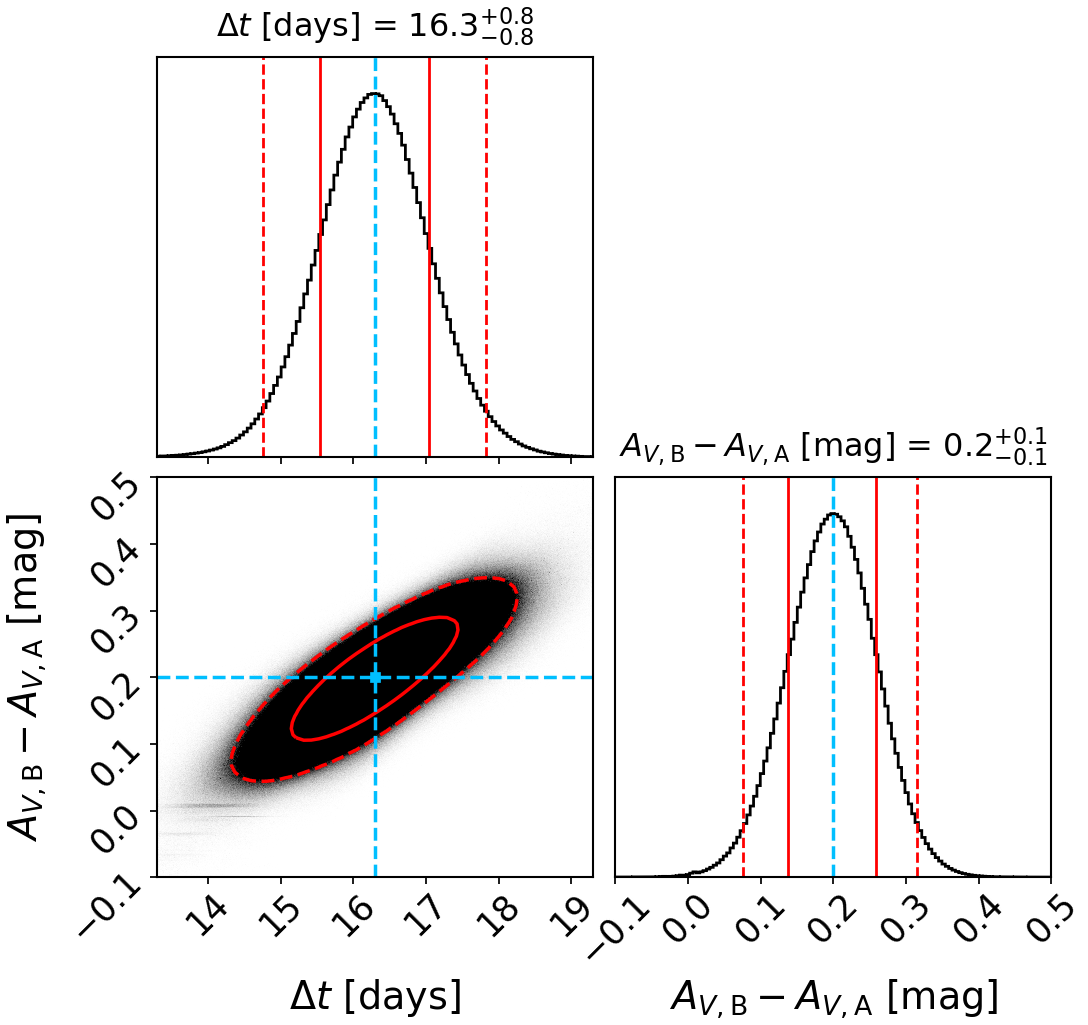}}\noindent\hspace*{1mm}%
\subfigure[Orders of fitting selected for case 2 with two color curves $u-g$ and $u-r$. We impose the same order for both color curves.]{\label{}\includegraphics[width=0.49\textwidth]{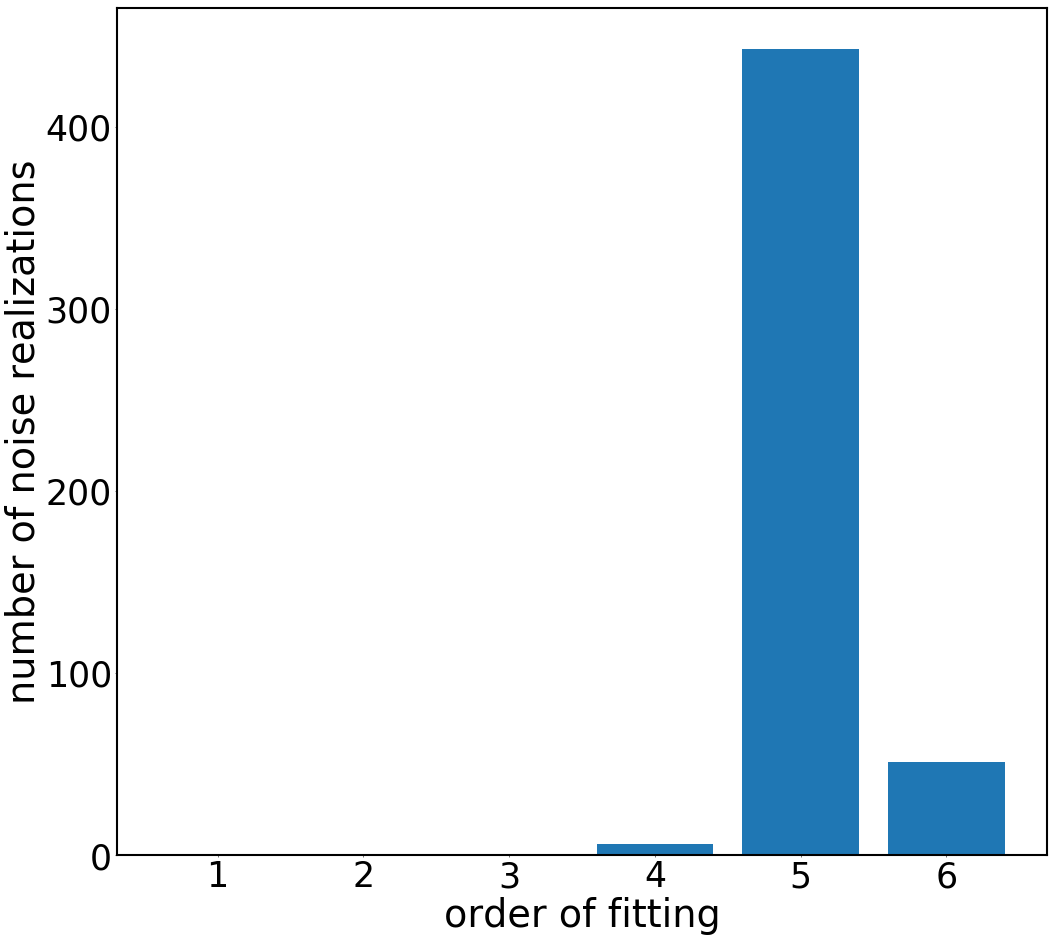}}\\\caption{\label{SN2005J_corner2}Same as Fig. \ref{SN2005J_corner1}, but for 500 noise realizations for case 2.}
\end{figure*}

\begin{figure*}[hbt!]
\centering
\subfigure[Case 3 with one color curve $u-g$.]{\label{}\includegraphics[width=0.49\textwidth]{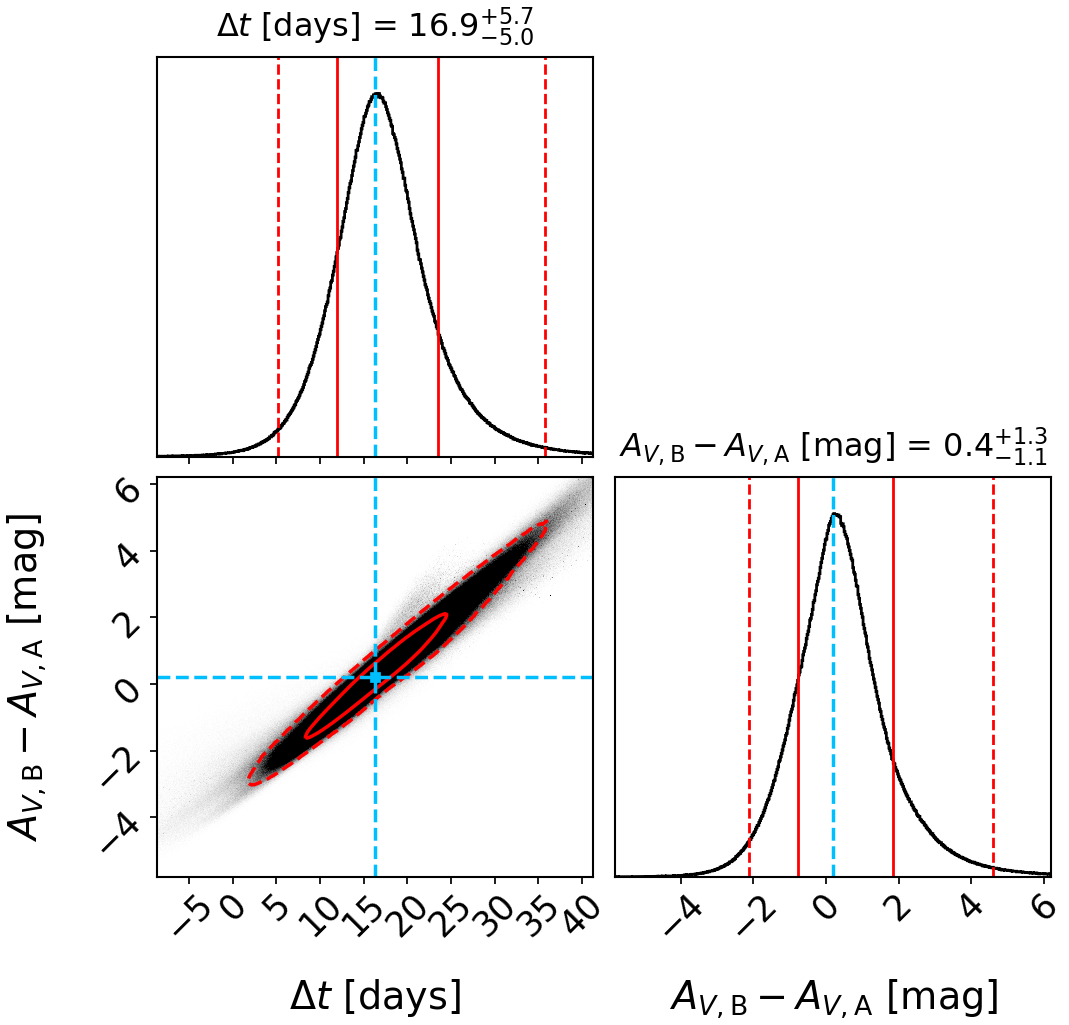}}\noindent\hspace*{1mm}%
\subfigure[Orders of fitting selected for case 3 with one color curve $u-g$.]{\label{}\includegraphics[width=0.49\textwidth]{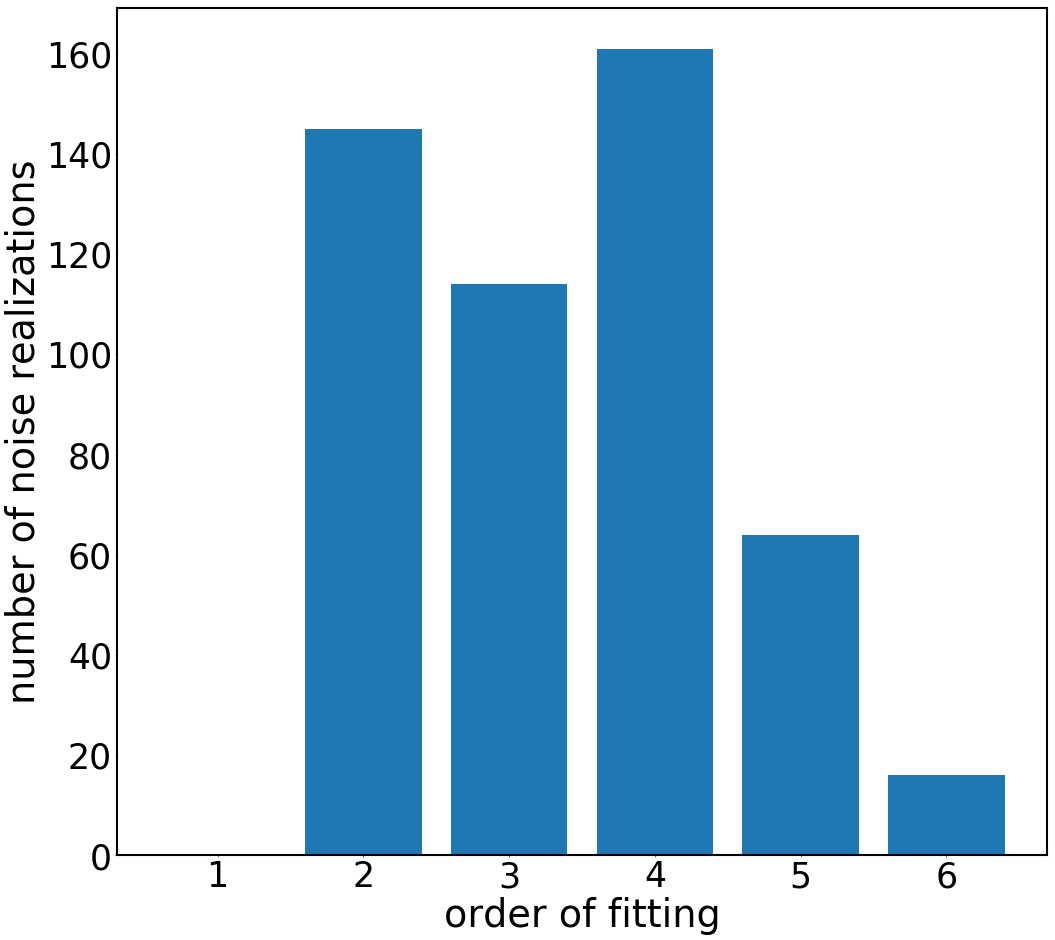}}\\
\subfigure[Case 3 with two color curves $u-g$ and $u-r$.]{\label{}\includegraphics[width=0.49\textwidth]{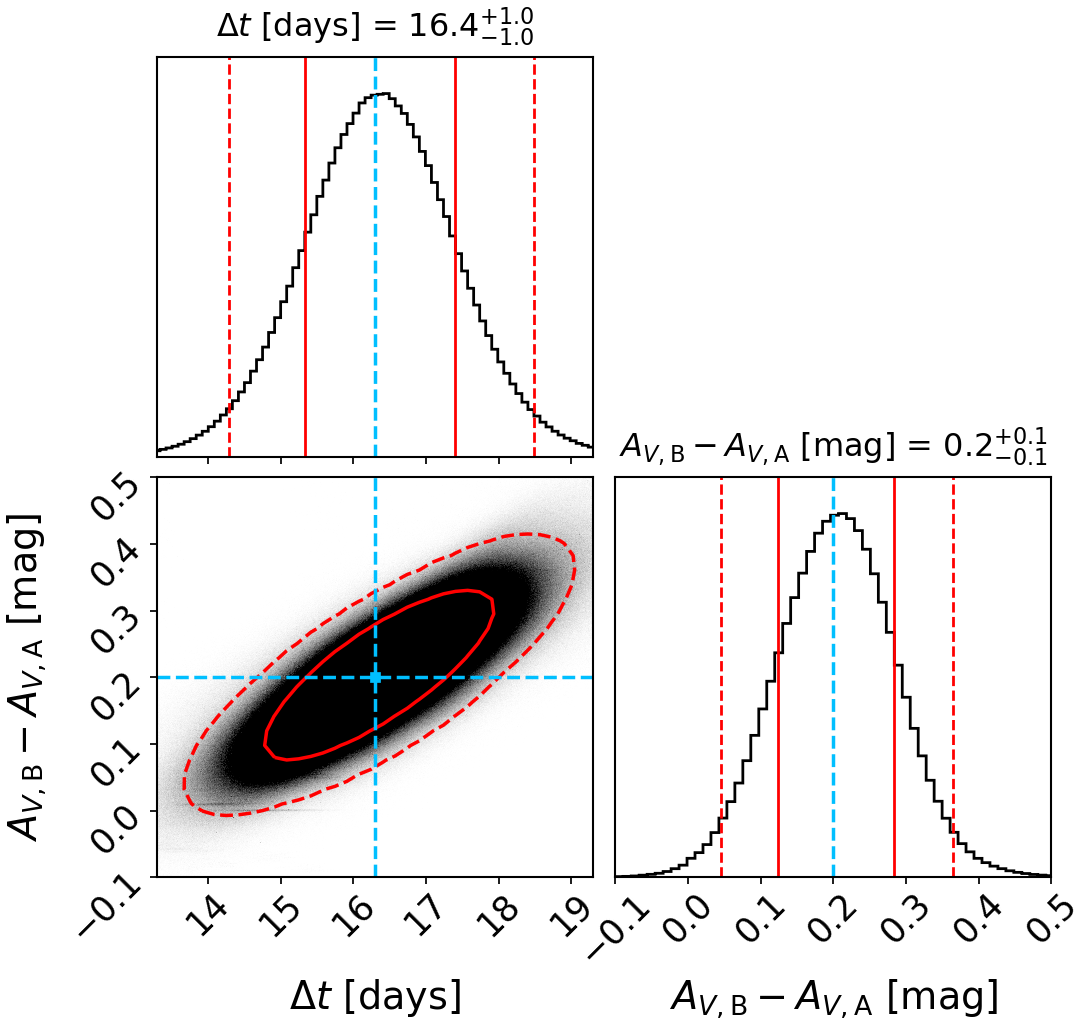}}\noindent\hspace*{1mm}%
\subfigure[Orders of fitting selected for case 2 with two color curves $u-g$ and $u-r$. We impose the same order for both color curves.]{\label{BIC_ugur_loss}\includegraphics[width=0.49\textwidth]{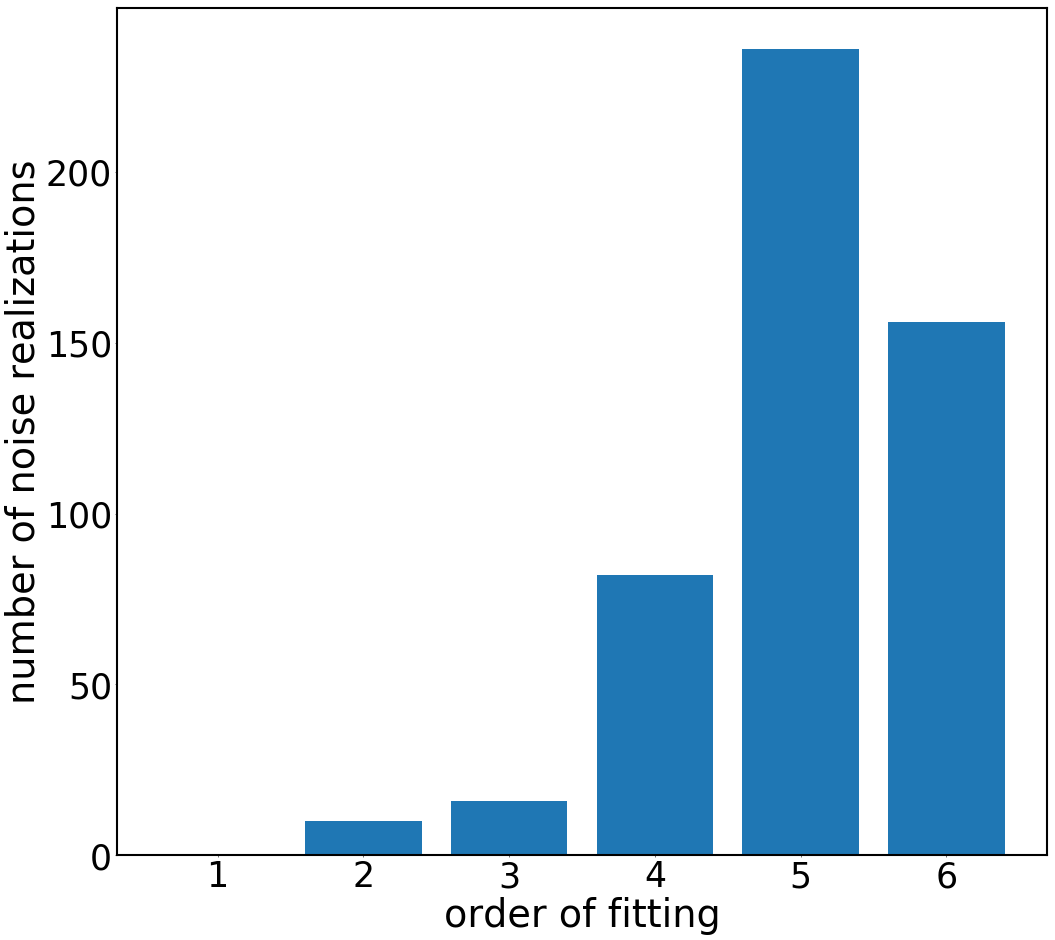}}\\\caption{\label{SN2005J_corner3}Same as Fig. \ref{SN2005J_corner1}, but for 500 noise realizations for case 3.}
\end{figure*}

The corner plots for mock data case 1 for one color curve $u-g$ and two color curves $u-g$ and $u-r$ are shown in Fig. \ref{SN2005J_corner1}. The cases 2 and 3 are shown in Fig. \ref{SN2005J_corner2} and Fig. \ref{SN2005J_corner3} respectively. In Fig. \ref{SN2005J_corner1} both corner plots share the same scale. The corner plots in Fig. \ref{SN2005J_corner2} and Fig. \ref{SN2005J_corner3} have different scaling, also compared to Fig. \ref{SN2005J_corner1}.
The time shifts and differential dust extinctions with uncertainties for all cases are additionally listed in Table \ref{table1}.

\begin{table*}[t]
\centering
\begin{tabular}{|m{3.2cm}||m{2.0cm}|m{1.8cm}||m{2.0cm}|m{1.8cm}||m{2.0cm}|m{1.8cm}|}
\hline 
$\Delta t_{\mathrm{true}} = 16.3 \ \mathrm{days}$ & \multicolumn{2}{c||}{}  & \multicolumn{2}{c||}{} & \multicolumn{2}{c|}{} \\ $(A_{V,\mathrm{B}}-A_{V,\mathrm{A}})_{\mathrm{true}} = 0.2 \ \mathrm{mag}$  &  \multicolumn{2}{c||}{\centering Case 1: high S/N} & \multicolumn{2}{c||}{\centering Case 2: low S/N} & \multicolumn{2}{c|}{\centering Case 3: low S/N, 35\% data loss} \\ & \multicolumn{2}{c||}{}  & \multicolumn{2}{c||}{} & \multicolumn{2}{c|}{}\\
\hline\hline  &  &  &  &  &  & \\
\centering $u-g$ & 16.30 $^{+0.14}_{-0.14}$ days & 0.20 $^{+0.02}_{-0.02}$ mag & 16.3 $^{+3.2}_{-3.6}$ days & 0.2 $^{+0.7}_{-0.8}$ mag & 16.9 $^{+5.7}_{-5.0}$ days & 0.4 $^{+1.3}_{-1.1}$ mag\\
 &  &  &  &  &  & \\
\hline  &  &  &  &  &  & \\
\centering $u-g$, $u-r$ & 16.29 $^{+0.08}_{-0.08}$ days & 0.20 $^{+0.01}_{-0.01}$ mag & 16.3 $^{+0.8}_{-0.8}$ days & 0.2 $^{+0.1}_{-0.1}$ mag & 16.4 $^{+1.0}_{-1.0}$ days & 0.2 $^{+0.1}_{-0.1}$ mag \\
 &  &  &  &  &  & \\
\hline 
\end{tabular} 
\caption{Results of the time-delay and differential dust extinction retrieval for each investigated S/N and color curve scenario. The time delays $\Delta t$ are listed on the left and the dust extinction $A_{V,\mathrm{B}}-A_{V,\mathrm{A}}$ on the right of each case.}
\label{table1}
\end{table*}

Considering case 1 for one and two color curves, we can retrieve the time shift and the differential extinction perfectly with relative uncertainties down to the 1\% level for typical delays and extinctions, which proves the functionality of the method. Furthermore, combining multiple color curves helps with constraining both the time delay and the differential dust extinction. For high S/N data in case 1, only high orders of the orthogonal polynomial fitting are selected via the BIC as shown in Figs. \ref{bar_1_1} and \ref{bar_1_2}.

When adjusting the S/N down for case 2, the result gets less precise, as expected. Considering only the $u-g$ color curve we cannot retrieve the time delay precisely enough to be of cosmology grade as defined in Sect. \ref{sec:intro}, unless the delays are longer than $\sim$ 35 days, to reach an uncertainty $\lesssim$ 10\% per lens system.
This gets even less precise for the most realistic case 3 including 35\% data loss, where uncertainties reach $\sim$ 5-6 days. $A_{V,\mathrm{B}}-A_{V,\mathrm{A}}$ cannot be constrained to better than $\sim$ 0.7 mag with a single color curve in low S/N cases.
By including the second color curve $u-r$, the constraint on the differential dust extinction is improved substantially to $<$ 0.1 mag in precision, which helps breaking the degeneracy between $\Delta t$ and $A_{V,\mathrm{B}}-A_{V,\mathrm{A}}$ and decreases the time delay uncertainty down to $\sim 1$ day.
For the lower S/N cases with one color curve, lower orders of fitting are selected via the BIC, because the higher uncertainties of the mock data leave more freedom in the functional form of the fit.
This gets shifted to higher orders again when a second color curve is taken into account, as it sets more constraints on the fit.

\section{Inference for non-zero redshift}
\label{sec: K_corr}

Up to now, we have investigated rest-frame photometry in the CSP filters $u,g,$ and $r$. In reality we would observe most lensed SNe II at redshifts around $z = 0.6$ \citep{Oguri2010}. Assuming a redshift of $z = 0.6$, the rest-frame $u,g,$ and $r$ bands would shift roughly into the observed $r,i,$ and $z$ bands respectively. In this section, we investigate how the shape of the kink is affected with the help of K-corrections and the rest-frame UV color curves of the Ultra Violet Optical Telescope (UVOT) on the Neil Gehrels Swift Observatory (SWIFT) \citep{Gehrels2004,SWIFT} SNe, which would be approximately observed in  the $u$ and $g$ band for $z = 0.6$.

\subsection{K-corrections}

\begin{figure}[hbt!]
\centering
\includegraphics[width=0.49\textwidth]{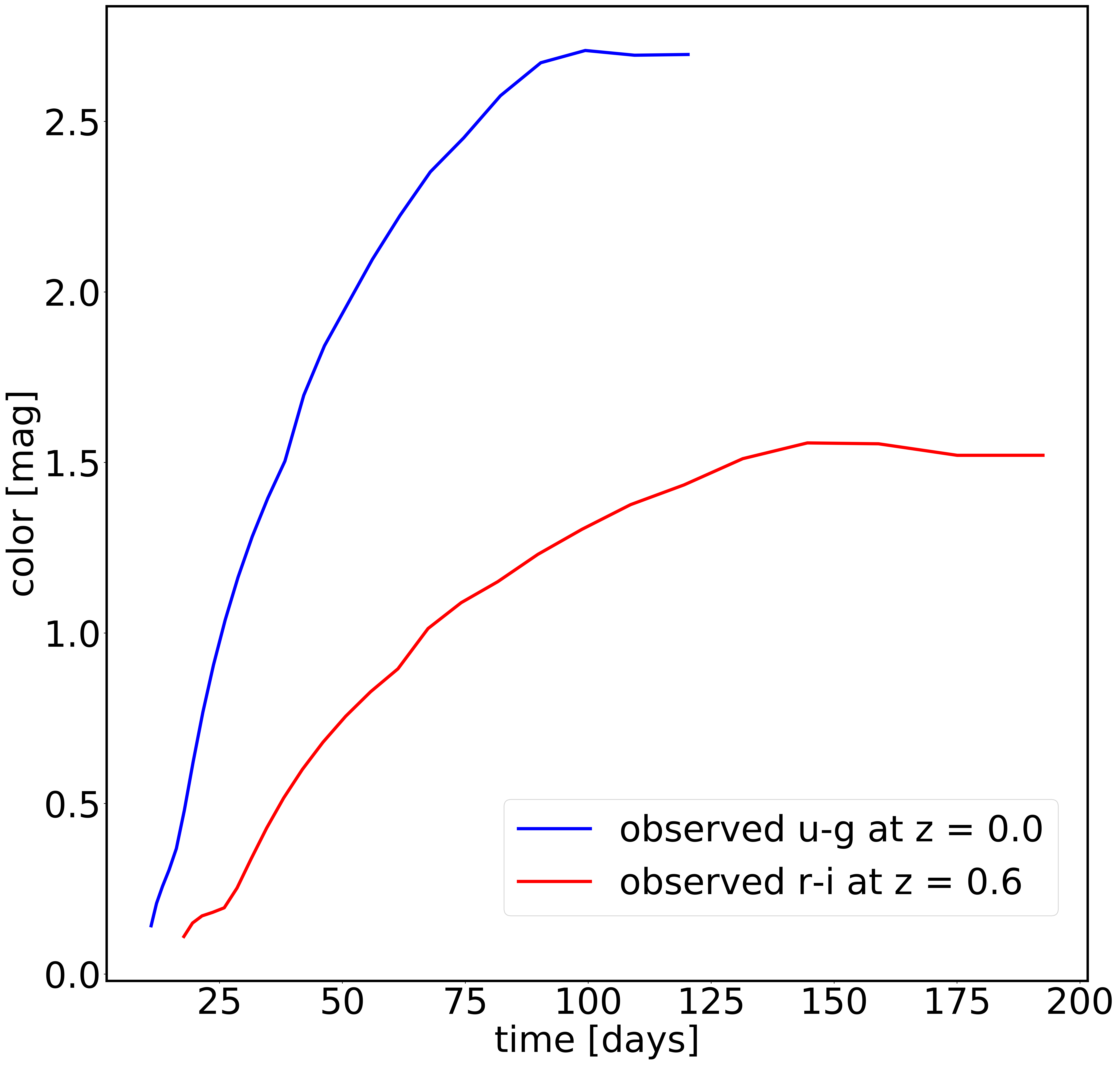}
\caption{\label{K_corr}Rest-frame $u-g$ color curve and corresponding redshifted $r-i$ color curve created from model spectra. The kink of the rest-frame color curve is slightly weakened by the K-correction.}
\end{figure}

For applying K-corrections, we need spectra, but for the CSP and SWIFT data we have no spectra available covering the rest-frame UV. We therefore use publicly available model spectra of SN IIP with temporal evolution created by \cite{Dessart2013}\footnote{\url{https://www-n.oca.eu/supernova/sn2p/ccsn_systematics.html}}.
We apply the K-correction to the photometry according to \cite{Hogg2002} by adding $K$ to the rest-frame magnitude in the rest-frame filter to retrieve the observed magnitude in the observed filter given the function:
\begin{equation}
\begin{array}{l}
K = -2.5 \log \left[\frac{1}{1+z} \frac{\int \lambda_{\mathrm{obs}} \ f(\lambda_{\mathrm{obs}}) \ R(\lambda_{\mathrm{obs}}) \ \mathrm{d}\lambda_{\mathrm{obs}}}{\int \lambda_{\mathrm{obs}} \ g^{R}(\lambda_{\mathrm{obs}}) \ R(\lambda_{\mathrm{obs}}) \ \mathrm{d}\lambda_{\mathrm{obs}}} \times \right. \\ 
\noindent\hspace*{28mm}%
\left. \frac{\int \lambda_{\mathrm{rest}} \ g^{Q}(\lambda_{\mathrm{rest}}) \  Q(\lambda_{\mathrm{rest}}) \ \mathrm{d}\lambda_{\mathrm{rest}}}{\int \lambda_{\mathrm{rest}} \ f((1+z) \lambda_{\mathrm{rest}}) \ Q(\lambda_{\mathrm{rest}}) \ \mathrm{d}\lambda_{\mathrm{rest}}} \right]
\end{array}
\end{equation}
with the observed wavelength $\lambda_{\mathrm{obs}}$, observed flux $f(\lambda_{\mathrm{obs}})$, transmission function of the filter in the observed frame $R(\lambda_{\mathrm{obs}})$, rest-frame wavelength $\lambda_{\mathrm{rest}}$, rest-frame flux $f(\lambda_{\mathrm{rest}})$, and transmission function of the filter in the rest-frame $Q(\lambda_{\mathrm{rest}})$.
$g^{R}(\lambda_{\mathrm{obs}})$ and $g^{Q}(\lambda_{\mathrm{rest}})$ are the spectral flux densities for a magnitude of zero in the AB system in the observed frame and the rest-frame, respectively.
We insert the LSST $u$ and $g$ bands for the rest-frame transmission functions and the $r$ and $i$ bands for the observed-frame transmission function, as $u$ and $g$ are roughly shifted into these bands for $z = 0.6$.
The resulting color curves are shown in Fig. \ref{K_corr}.
The strength of the kink is defined by the change of the slopes of the color curve before the kink and after the kink. The stronger the change, the stronger the kink.
Applying the K-correction, the kink of the color curve, which occurs in this case around day 95 is slightly weakened. As this change is not too drastic, we expect that redshifted SNe IIP color curves still exhibit a sufficiently strong kink for our method to work, especially when combining multiple color curves.
To include redshifted color curves using the observed $u$ and $g$ bands, we investigate next the evolution of rest-frame UV color curves.

\subsection{Rest-frame UV color curves}
\label{sec: SWIFT}

\begin{figure}[hbt!]
\centering
\includegraphics[width=0.49\textwidth]{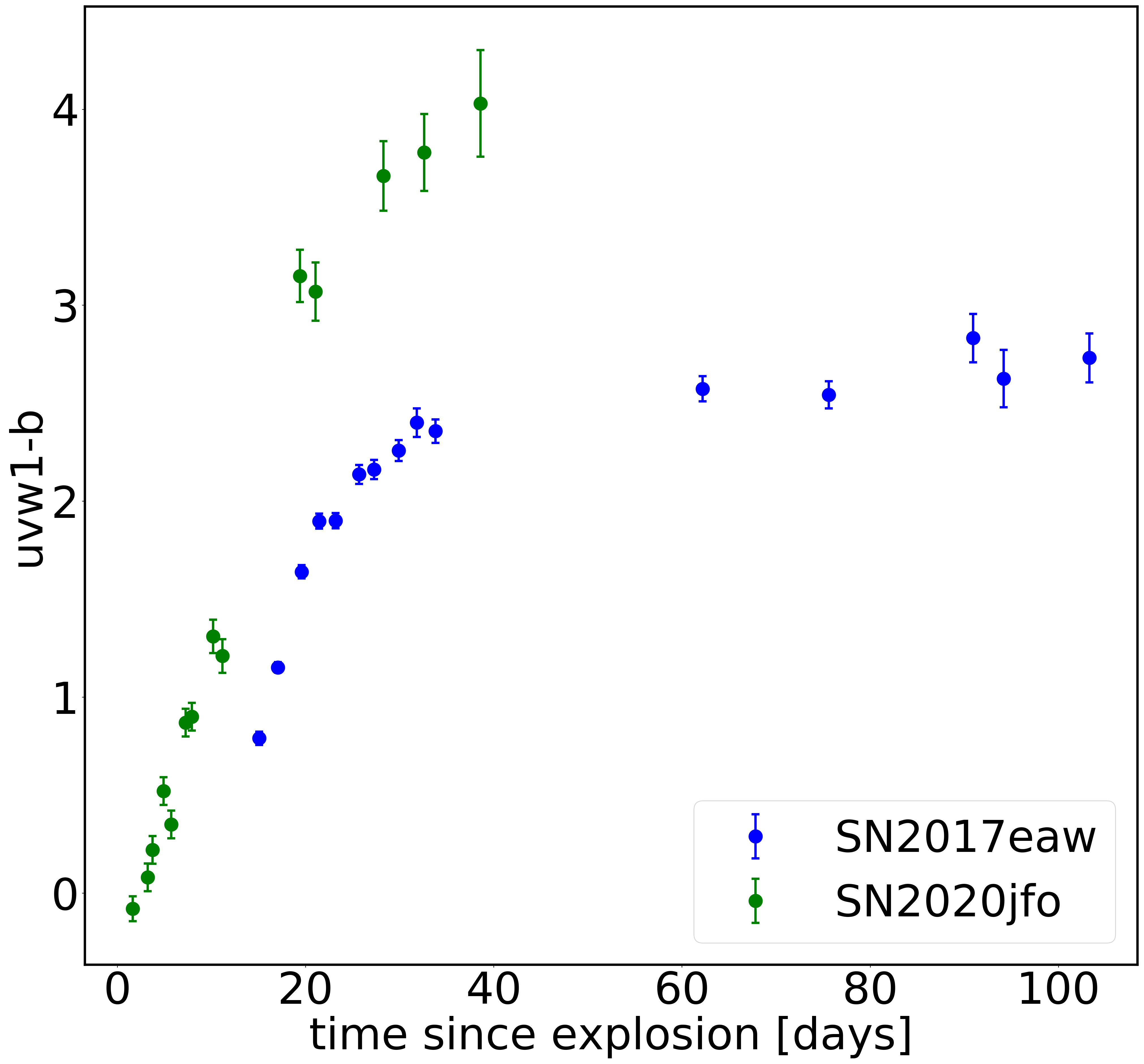}
\caption{\label{color_curve_mock_SWIFT}$uvw1-b$ color curves of SWIFT SNe SN2017eaw (blue) \citep{Szalai2019} and SN2020jfo (green) \citep{Sollerman2021}.}
\end{figure}

\begin{figure}[hbt!]
\centering
\includegraphics[width=0.49\textwidth]{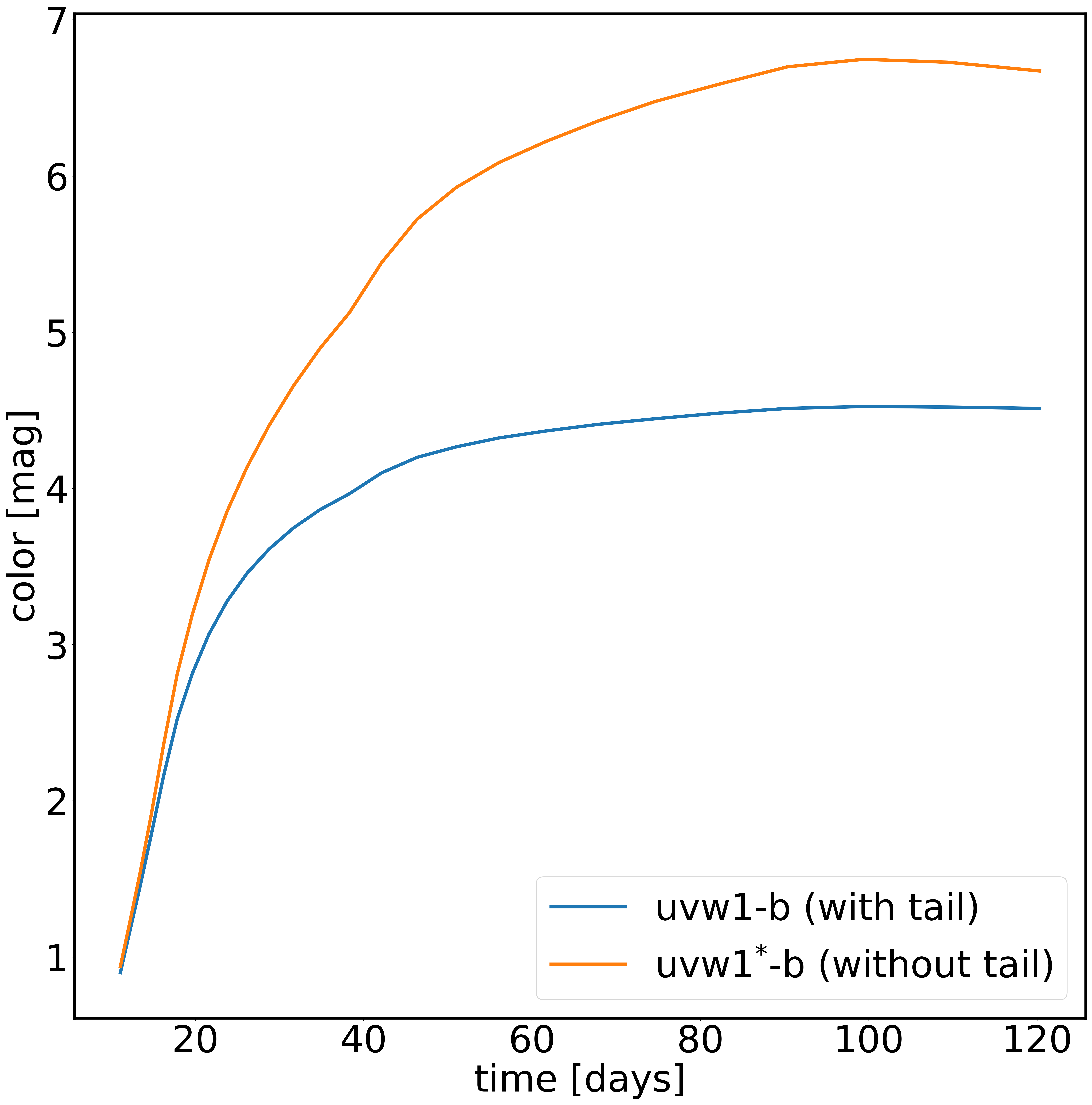}
\caption{\label{uv_tail_plot}Color curves $uvw1-b$ (blue) and $uvw1^{*}-b$ (orange) created with the \cite{Dessart2013} model spectra using the SWIFT filters $uvw1$, $b$, and the modified $uvw1^{*}$, where the red tail has been cut. The color curve $uvw1^{*}-b$ still shows a sufficiently strong kink for our time delay method to work.}
\end{figure}

To estimate how rest-frame UV color curves are redshifted into the range of LSST bands we investigate the UV color curves with the UVOT filters $uvw1$ and $b$ of the two SWIFT SNe SN2020jfo and SN2017eaw. We have to limit the SWIFT light curves to the UVOT rest-frame $uvw1$ and $b$ bands, as only these two bands have a good coverage of the epochs around the kink observed in the color curves. The data for SN2020jfo is taken from \cite{Sollerman2021} and for SN2017eaw from \cite{Szalai2019}.
The rest-frame $uvw1-b$ color curves of both SNe in Fig. \ref{color_curve_mock_SWIFT} show a particularly strong kink. We find that this is a result of the long wavelength tail of the $uvw1$ filter reaching into the visible spectral range. 
We investigate how cutting this $uvw1$ filter tail away at $\sim$3347.5 $\AA$ to create a filter $uvw1^{*}$ affects the kink in color curves, by creating artificial color curves $uvw1-b$ and $uvw1^{*}-b$ with the spectra by \cite{Dessart2013}. The resulting color curves are shown in Fig. \ref{uv_tail_plot}.
$uvw1^{*}-b$ shows a weaker kink than $uvw1-b$, but still stronger than the $u-g$ rest-frame color curve from Fig. \ref{K_corr}. We therefore conclude that bluer colors tend to show even stronger kinks for SNe IIP, which helps us constraining the time delay for redshifted SNe IIP.

\section{Discussion and conclusion}
\label{sec: Discussion_Conclusion}

We have developed a Metropolis Hastings MCMC based method of fitting lensed SN IIP color curves with orthogonal polynomial functions to retrieve the time delay $\Delta t$ and the differential dust extinction $A_{V,\mathrm{B}}-A_{V,\mathrm{A}}$ between two lensing images A and B.

In the high-S/N limit, the method works flawlessly using either one or two color curves, and the time delay $\Delta t$ and the differential dust extinction $A_{V,\mathrm{B}}-A_{V,\mathrm{A}}$ can be retrieved precisely without bias. We note that $A_{V}$ of each individual image is only determined precisely for the case of one color curve assuming the knowledge of the macro-magnification.
Orthogonal polynomial functions are flexible enough to fit data which may follow different functional forms.

Considering more realistic cases with low S/N and data loss because of observational constraints, the inference of $\Delta t$ and especially with $A_{V,\mathrm{B}}-A_{V,\mathrm{A}}$ is more uncertain. For one color curve, the precision of the time delay degrades to $\gtrsim$ 3 days. Adding the information of a third light curve band helps with constraining both parameters. 
Fitting two color curves calculated from light curves in 3 bands gives more constraints on $A_{V,\mathrm{B}}-A_{V,\mathrm{A}}$ helping to break the degeneracy between the time delay and the differential dust extinction. This has also the advantage that absolute magnification cancels out between the two color curves assuming we assume the correct $R_{V}$ and extinction law law during sampling.

We are aware that our assumption of knowing perfectly $R_{V}$ and the correct extinction law is strong and simplifies the determination of $\Delta t$ and $A_{V,\mathrm{B}}-A_{V,\mathrm{A}}$. 
We show in Appendix \ref{RV} that for combining two realistic mock color curves with low S/N and data loss using a different extinction law, $\Delta t$ can still be determined precisely and without bias, which is the main objective of this work. The accuracy of $A_{V,\mathrm{B}}-A_{V,\mathrm{A}}$ is depending more on $R_{V}$ than on the specific extinction law in the wavelength range of the LSST filters.

Despite all measures we take to create realistic mock color curves, we anticipate some additional uncertainties in real data. Three additional nuisance effects may be expected for real data: the decrease in S/N because of the additional shot noise from the host galaxy of the SN and the lens, photometric errors due to the blending of the SN images, and the impact of the seeing on the above two effects.
The influence of these effects will have to be studied in the future when more observed lensed SNe IIP are available.
Another simplification of the existing framework is the use of a non-redshifted reddening. This should not impact our results heavily within the obtained uncertainties, as the observed bands probe a region of the extinction laws that is monotonically decreasing and the lens will be likely at low redshifts around $z$ = 0.2 to $z$ = 0.4 \citep{Oguri2010}.

Considering all rest-frame cases presented in Sect. \ref{sec:results} as well as Appendix \ref{App:diff_model}, \ref{lin_case}, and \ref{diff_shift}, we can determine several trends:
\begin{itemize}
	\item The more color curves we take into account during sampling the better we can determine $A_{V,\mathrm{B}}-A_{V,\mathrm{A}}$, which results also in a better constraint on the time delay, as different color curves can help to break the degeneracy in $\Delta t$ and $A_{V,\mathrm{B}}-A_{V,\mathrm{A}}$.
	\item The data needed for this method to work at its best should ideally be a dedicated follow up multiband observation with a cadence of not more than two rest-frame days, and a depth that is 1 mag deeper than LSST to have a high enough S/N for typical lensed-SN redshifts and magnifications \citep{Huber2022}.
	\item The K-correction investigation in Sect. \ref{sec: K_corr} of rest-frame UV bands redshifted into the LSST bands $u$, $g$, $r$, and $i$, shows strong enough kinks to be usable for our MCMC retrieval technique. Bluer colors tend to show stronger kinks for SNe IIP, hence the redshift of a lensed SN IIP helps us to get better time-delay measurements with ground-based photometry.
	\item A higher S/N ratio, a larger number of considered color curves, and lower data loss result in a higher selected order of fitting for the orthogonal polynomials as the data impose higher constraints.
\end{itemize}

The method is precisely enough for obtaining cosmology grade time delays for gravitationally lensed SNe IIP with realistic uncertainties around $\sim$ 1.0 day. For 20 SNe IIP with time delays longer than 20 days we will thus be able to constrain $H_{0}$ with an uncertainty of $\sim$ 1\% during the era of LSST \citep{Suyu2020}.
Comparing the results of this work to our previous paper HOLISMOKES V \citep{Bayer2021}, which used a combination of absorption lines within spectra of LSNe IIP, we achieve a similar accuracy and precision of $\sim$ 1.0 day.

\newpage

\begin{acknowledgements}

We thank F.~Dux, M.~Millon, and F.~Courbin for discussions on realistic data loss in observations, and J.~Anderson and T.~de Jaeger for providing us the observational data of SN2005J and other SNe of the CSP to develop and test our code, and E. Fitzpatrick for discussions on dust extinction laws. We acknowledge the use of public data from the Swift data archive.
We thank the Max Planck Society for support through the Max Planck
Research Group and Fellowship for SHS.
This project has received funding from the
European Research Council (ERC) under the European Union’s Horizon
2020 research and innovation programme (LENSNOVA: grant agreement
No. 771776; COSMICLENS: grant agreement No. 787886).
This research is supported in part by the Excellence Cluster ORIGINS which is funded by the Deutsche Forschungsgemeinschaft (DFG, German Research Foundation) under Germany's Excellence Strategy -- EXC-2094 -- 390783311.
\end{acknowledgements}

\bibliographystyle{aa}
\bibliography{Color_curve_time_delay_retrieval.bib}

\FloatBarrier

\appendix

\section{Fitting with a different model}
\label{App:diff_model}

For our main study we used the same functional form (orthogonal polynomials) to generate and fit to the mock data. In this appendix, we explore if the method might produce model induced biased results  if the mock data is created with a spline model and sampled with orthogonal polynomials. Further, we do not expect real data to follow the behavior of orthogonal polynomials, nor a spline, and we want to ensure that the model used during sampling is flexible enough to fit to the data based on a different model. 

For the mock data creation we create a spline function consisting of two third-order polynomials connected at one knot $t_{\mathrm{knot}}$ defined as:
 \begin{equation}\label{equ:spline}
C(\vec{\eta},t)=\begin{cases}
C_{1}(\vec{\eta}_{1},t) = c_{1} t^{3} + c_{2} t^{2}\\
 \ \ \ \ \ \ \ \ \ \ \ \ \ \ \ \ \ \ + c_{3} t + c_{4}  \ \ \ \ \ \ \ \ \ \ \ \ \ \ \ \ \ \ \text{for} \ t \leq t_{\mathrm{knot}}\\
C_{2}(\vec{\eta},t) \ \ = c_{4} + c_{5} t^{3} \\
 \ \ \ \ \ \ \ \ \ \ \ \ \ \ \ \ \ \ + c_{6} t^{2} + c_{7} t \\ 
 \ \ \ \ \ \ \ \ \ \ \ \ \ \ \ \ \ \ + (c_{1}-c_{5}) t_{\mathrm{knot}}^{3} \\ 
 \ \ \ \ \ \ \ \ \ \ \ \ \ \ \ \ \ \ + (c_{2}-c_{6}) t_{\mathrm{knot}}^{2} \\
  \ \ \ \ \ \ \ \ \ \ \ \ \ \ \ \ \ \ + (c_{3}-c_{7}) t_{\mathrm{knot}} \ \ \ \ \ \ \ \ \ \ \text{for} \ t > t_{\mathrm{knot}}
\end{cases}
\end{equation}
with the additional condition of the same slope at the knot:
\begin{equation}\label{equ:slope}
\frac{\partial C_{1}(\vec{\eta}_{1},t_{\mathrm{knot}})}{\partial t} = \frac{\partial C_{2}(\vec{\eta},t_{\mathrm{knot}})}{\partial t}
\end{equation}
The collective parameter vectors are defined as $\vec{\eta} = (c_{1},c_{2},c_{3},c_{4},c_{5},c_{6},c_{7},t_{\mathrm{knot}})$ and $\vec{\eta}_{1} = (c_{1},c_{2},c_{3},c_{4})$ for $C_{1}(\eta_{1},t)$. With the slope condition, we can eliminate the parameter $c_{7}$ and express it via the other parameters:
\begin{equation}\label{equ:c7}
c_{7} = 3t_{\mathrm{knot}}^{2}(c_{1}-c_{5}) + 2t_{\mathrm{knot}}(c_{2}-c_{6}) + c_{3}.
\end{equation}
We insert Eq. \ref{equ:c7} into Eq. \ref{equ:spline}:
 \begin{equation}
C(\vec{\eta},t)=\begin{cases}
C_{1}(\vec{\eta}_{1},t) = c_{1} t^{3} + c_{2} t^{2}\\
 \ \ \ \ \ \ \ \ \ \ \ \ \ \ \ \ \ \ + c_{3} t + c_{4}  \ \ \ \ \ \ \ \ \ \ \ \ \ \ \ \ \ \ \ \text{for} \ t \leq t_{\mathrm{knot}}\\
C_{2}(\vec{\eta},t) \ \ = c_{5} t^{3} + c_{6} t^{2} \\
 \ \ \ \ \ \ \ \ \ \ \ \ \ \ \ \ \ \ + [3t_{\mathrm{knot}}^{2}(c_{1}-c_{5}) \\
 \ \ \ \ \ \ \ \ \ \ \ \ \ \ \ \ \ \ + 2t_{\mathrm{knot}}(c_{2}-c_{6})\\
 \ \ \ \ \ \ \ \ \ \ \ \ \ \ \ \ \ \ + c_{3}] t + (c_{1}-c_{5}) t_{\mathrm{knot}}^{3}\\
 \ \ \ \ \ \ \ \ \ \ \ \ \ \ \ \ \ \ + (c_{2}-c_{6}) t_{\mathrm{knot}}^{2} \\ 
 \ \ \ \ \ \ \ \ \ \ \ \ \ \ \ \ \ \ + (c_{3}-[3t_{\mathrm{knot}}^{2}(c_{1}-c_{5})\\ 
 \ \ \ \ \ \ \ \ \ \ \ \ \ \ \ \ \ \ + 2t_{\mathrm{knot}}(c_{2}-c_{6})\\
 \ \ \ \ \ \ \ \ \ \ \ \ \ \ \ \ \ \ + c_{3}]) t_{\mathrm{knot}} + c_{4} \ \ \ \ \ \ \ \ \ \ \ \text{for} \ t > t_{\mathrm{knot}} \\
\end{cases}
\end{equation}
and redefine the parameter vector $\vec{\eta} = (c_{1},c_{2},c_{3},c_{4},c_{5},c_{6},t_{\mathrm{knot}})$.
Besides the change in generating mock data, we follow the same approach as described in Sect. \ref{sec:MCMC}.

\begin{figure}[hbt!]
\centering
\includegraphics[width=0.49\textwidth]{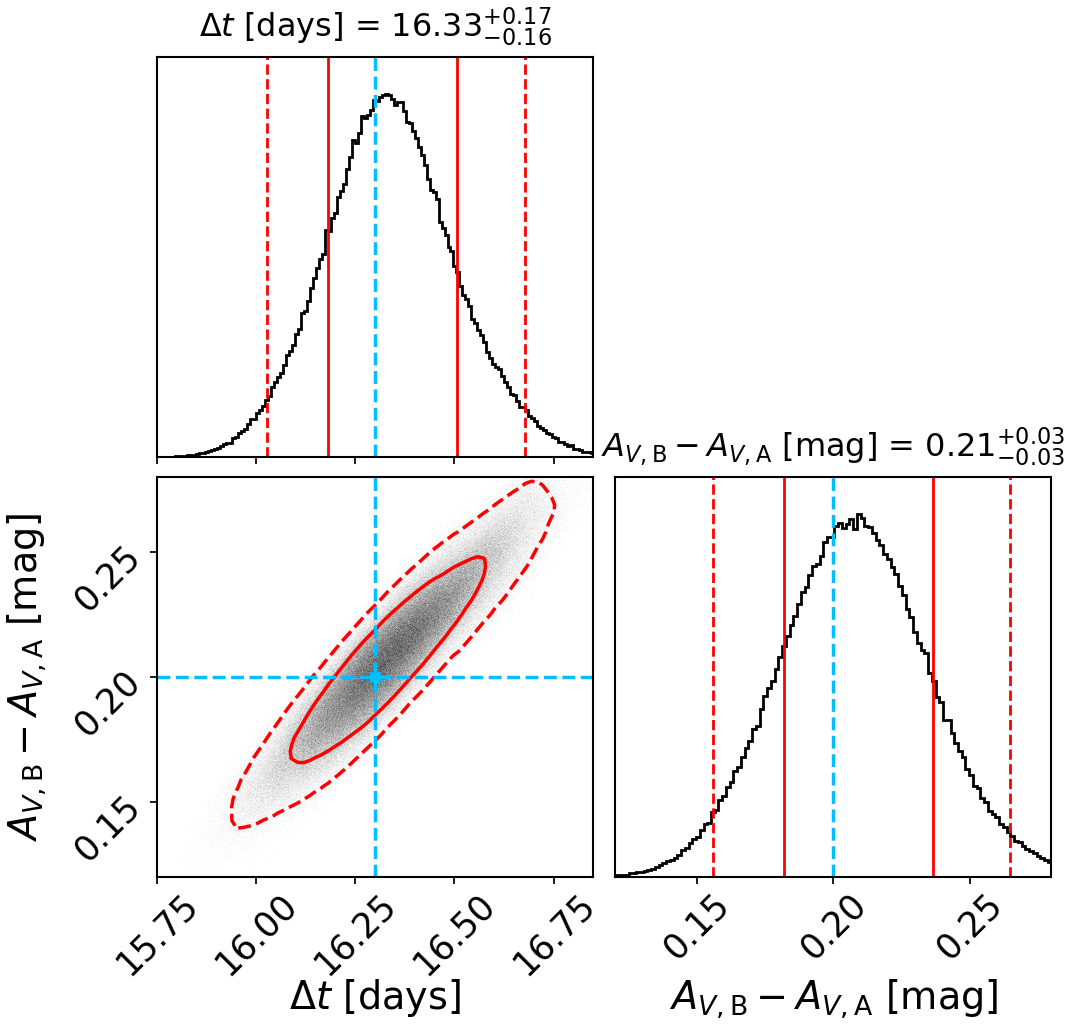}
\caption{\label{spline_mock_corner}Corner plot for mock data created with spline fitting including 100 noise realizations with high S/N. The initial input values shown in light blue are $\Delta t$ = 16.3 days and $A_{V,\mathrm{B}}-A_{V,\mathrm{A}}$ = 0.2 mag.}
\end{figure}

\begin{figure}[hbt!]
\centering
\includegraphics[width=0.49\textwidth]{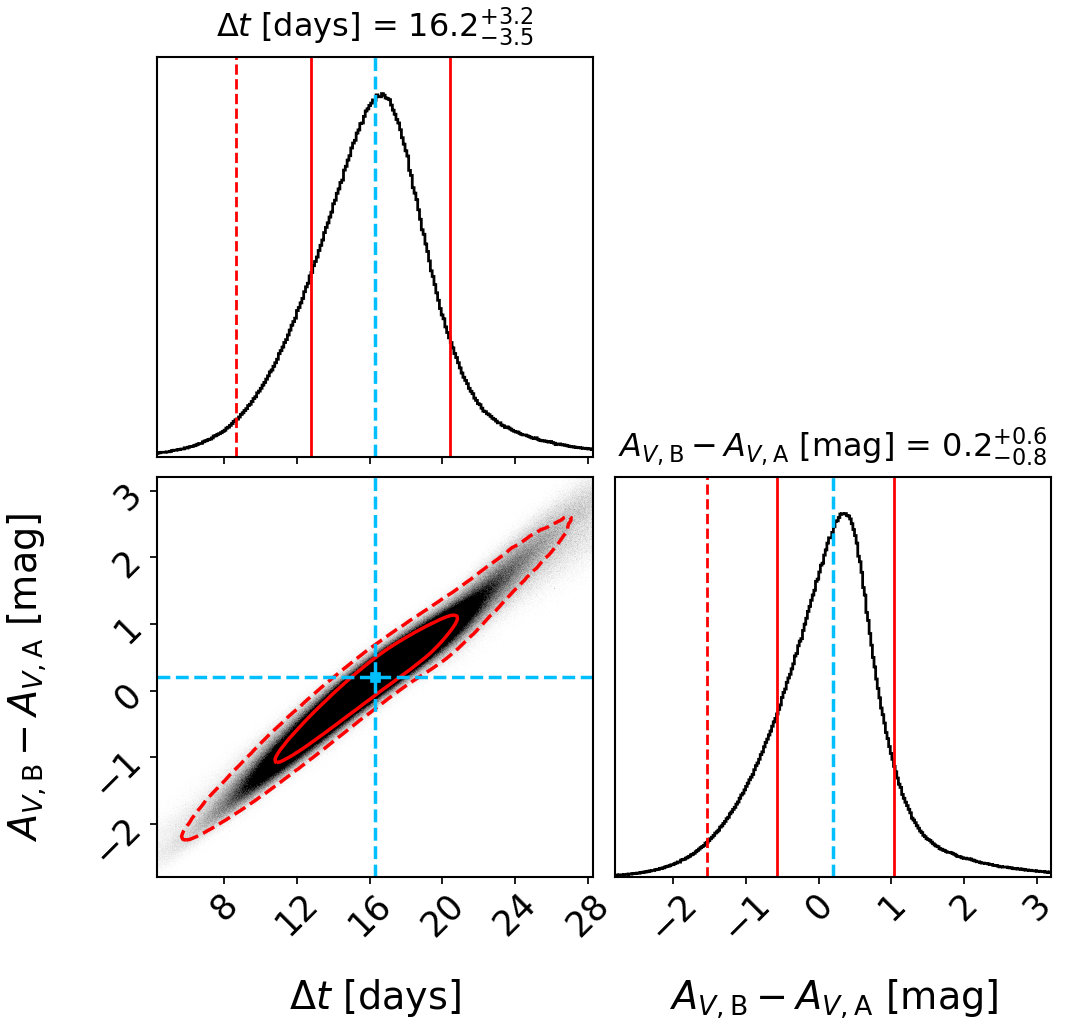}\caption{\label{spline_mock_corner_large}Corner plot for mock data created with spline fitting including 500 noise realizations with low S/N. The initial input values shown in light blue are $\Delta t$ = 16.3 days and $A_{V,\mathrm{B}}-A_{V,\mathrm{A}}$ = 0.2 mag.}
\end{figure}

\begin{figure}[hbt!]
\centering
\includegraphics[width=0.49\textwidth]{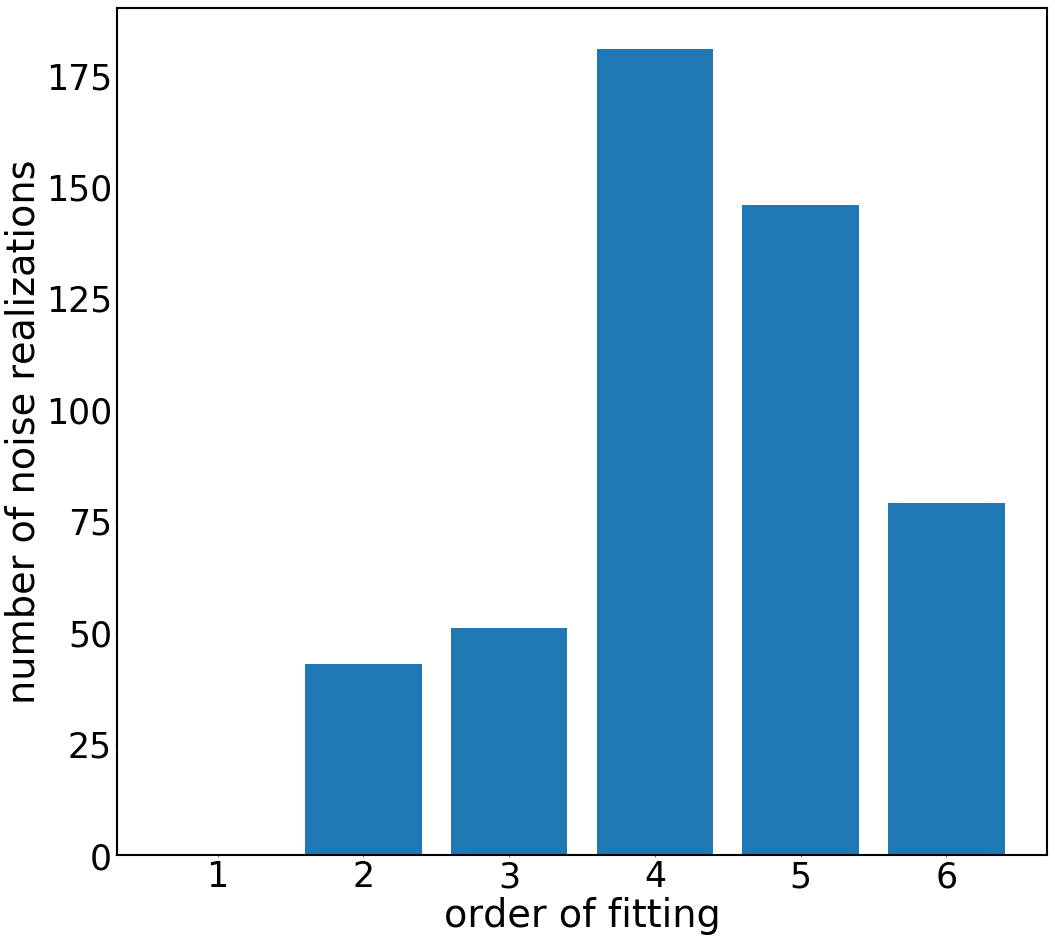}\caption{\label{spline_mock_corner_large_orders}Histogram of the selected orthogonal polynomial orders of fitting using the BIC for the low S/N mock data created with spline fitting.}
\end{figure}

We first investigate high S/N with only one color curve $u-g$ and 100 noise realizations as this simple case will show most easily if a bias occurs.
The resulting corner plot is shown in Fig. \ref{spline_mock_corner} for which only the 6th order was selected by the BIC. It has a slight bias in the time shift and the differential extinction, although the bias in $\Delta t$ is only 0.03 days which is subpercent for most lensed SN systems of interest in our program.
We further check the case of low S/N with one color curve $u-g$ and 500 noise realizations.
Because of the low S/N the result shown in Fig. \ref{spline_mock_corner_large} is unbiased within the uncertainties. The selected orders are shown in the bar plot in \ref{spline_mock_corner_large_orders}
This proves that for the realistic cases with low S/N we can retrieve unbiased results even with different models used in mock data creation and sampling.

\section{Completely Degenerate Case}
\label{lin_case}

\begin{figure}[hbt!]
\centering
\includegraphics[width=0.49\textwidth]{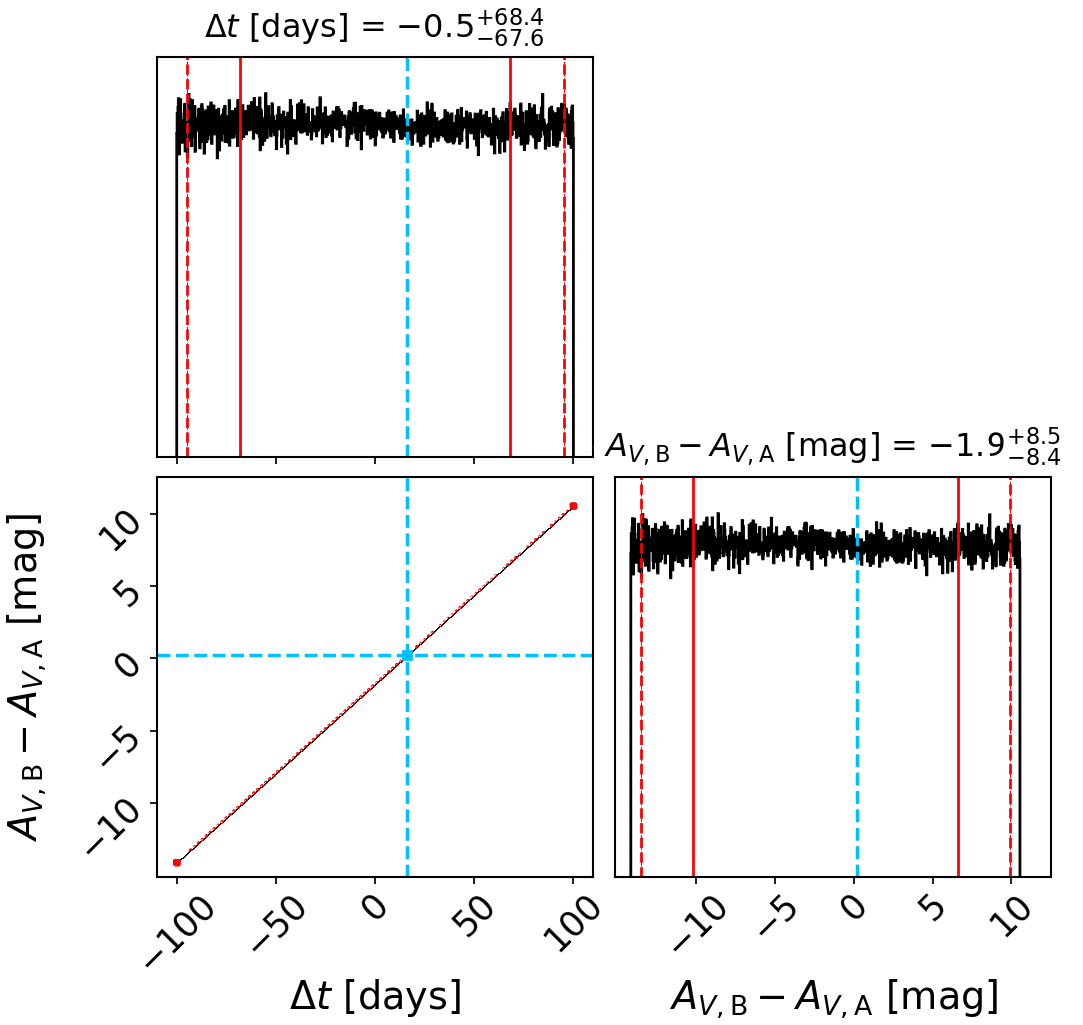}\caption{\label{degenerate_corner}Corner plot combining 100 noise realizations of perfect linear mock color curves $u-g$. Because of the linearity of the mock data, a complete degeneracy is retrieved.}
\end{figure}

Another test of the functionality of our code is to create a mock color curve case which is perfectly linear in both images and try to retrieve a time delay with our method. We expect that the preferred order selected with the BIC of the orthogonal polynomials is one and the distribution in the time shift and differential extinction is completely degenerate. As it can be seen in Fig. \ref{degenerate_corner}, we can retrieve an exact degeneracy up to the borders of the prior and during the BIC selection only order one was selected. Even if $A_{V,\mathrm{B}}-A_{V,\mathrm{A}}$ has no prior boundaries as it has an unbound uniform prior, it is limited by the boundaries of the prior for $\Delta t$. This test shows that for featureless (linear) color curves, our color curve time-delay determination breaks down.

\section{Different shifts}
\label{diff_shift}

\begin{figure}[hbt!]
\centering
\includegraphics[width=0.49\textwidth]{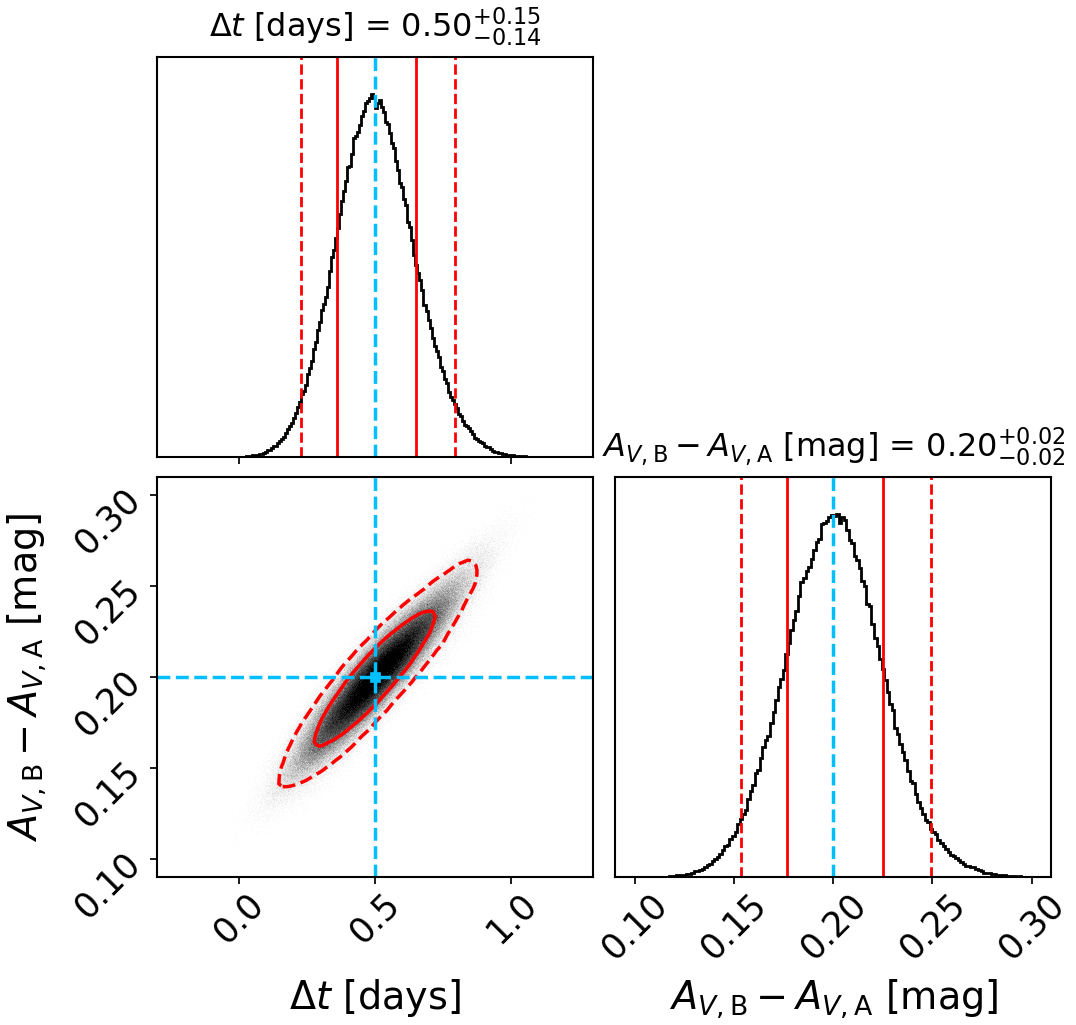}
\caption{\label{small_shift}Corner plot of 100 noise realizations of high S/N mock data with $\Delta t_{\mathrm{small}}$ = 0.5 days and $A_{V,\mathrm{B}}-A_{V,\mathrm{A}}$ = 0.2 mag}
\end{figure}

\begin{figure}[hbt!]
\centering
\includegraphics[width=0.49\textwidth]{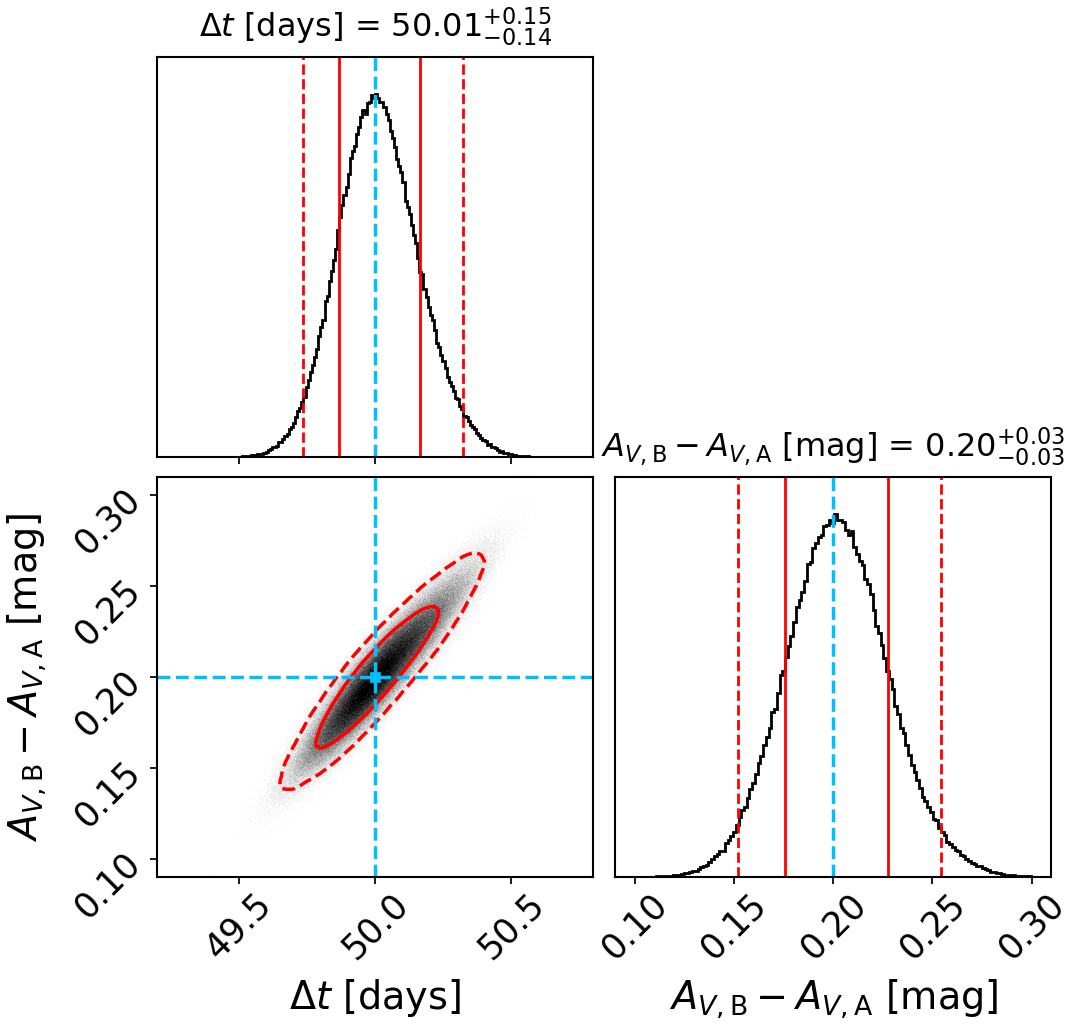}
\caption{\label{large_shift}Corner plot of 100 noise realizations of high S/N mock data with $\Delta t_{\mathrm{large}}$ = 50.0 days and $A_{V,\mathrm{B}}-A_{V,\mathrm{A}}$ = 0.2 mag}
\end{figure}

Our main study focused only on one time shift $\Delta t$ = 16.3 days. Here we also investigate two cases with $\Delta t_{\mathrm{small}}$ = 0.5 days and $\Delta t_{\mathrm{large}}$ = 50 days, to check that any low or high time shift can be retrieved. We do not explore time shifts longer than 50 days, as these can be also detected by eye and shifted manually into the range of $\Delta t < $ 50 days, which ensures that our code can be applied to any time shift.
To create the two plots in Fig. \ref{small_shift} and Fig. \ref{large_shift} we follow the same procedure as explained in Sect. \ref{sec:MCMC} with 100 noise realizations and high S/N data, just switching out the time shift in the mock data creation to $\Delta t_{\mathrm{small}}$ = 0.5 days and $\Delta t_{\mathrm{large}}$ = 50 days. For both new time shifts a similar distribution of the fitting orders as for case 1 with one color curve shown in Fig. \ref{bar_1_1} is selected.
For both cases we can retrieve the time delay and differential dust extinction as well as for case 1 considered in Sect. \ref{sec:results} using only one color curve $u-g$, which confirms that the code can be applied to any time shift within 0.5 days to 50 days.

Besides all the testing presented here in the appendix, we also used the nested sampling Monte Carlo algorithm MLFriends \citep{Buchner2014,Buchner2019} from the UltraNest\footnote{\url{https://johannesbuchner.github.io/UltraNest/}} package \citep{Buchner2021} with our technique and obtained the same results, confirming that our sampling code works correctly. Using our own code gave us even faster results, because we do not calculate the Bayesian evidence.

\section{Different extinction laws for mock data and sampling}
\label{RV}

In the main study, we know and adopt the correct extinction law to sample the color curves. In reality this will not be the case and we might assume $R_{V}$ wrongly or use an extinction law not fitting the real data. Therefore, we test three additional cases with 100 noise realizations for the realistic case of low S/N with 35\% data loss for two color curves $u-g$ and $u-r$. All three cases are still based on the same mock data as explained in Sect. \ref{sec:color_curves}.

The first test case T1 sets $R_{V} = $ 2.0 for the sampling. 
The second case T2 uses $R_{V} = $ 5.0 for sampling. For both T1 and T2 we still use Cardelli extinction laws in the sampling process described in Sect. \ref{sec:MCMC}.
The third test case T3 applies more recent extinction laws from \cite{Gordon2023}, referred to as Gordon laws, in the sampling, but keeping $R_{V} = $ 3.1. The Gordon laws have a similar approach as the Cardelli laws in Eq. \ref{eq: Cardelli}, but are based on more recent measurements for $a(x)$ and $b(x)$ and are normalized differently:
\begin{equation} \label{eq: Gordon}
\frac{A_{\lambda}}{A_{V}} = a(x) + \frac{b(x)}{R_{V}} - \frac{b(x)}{3.1}.
\end{equation}

\begin{figure}[hbt!]
\centering
\includegraphics[width=0.49\textwidth]{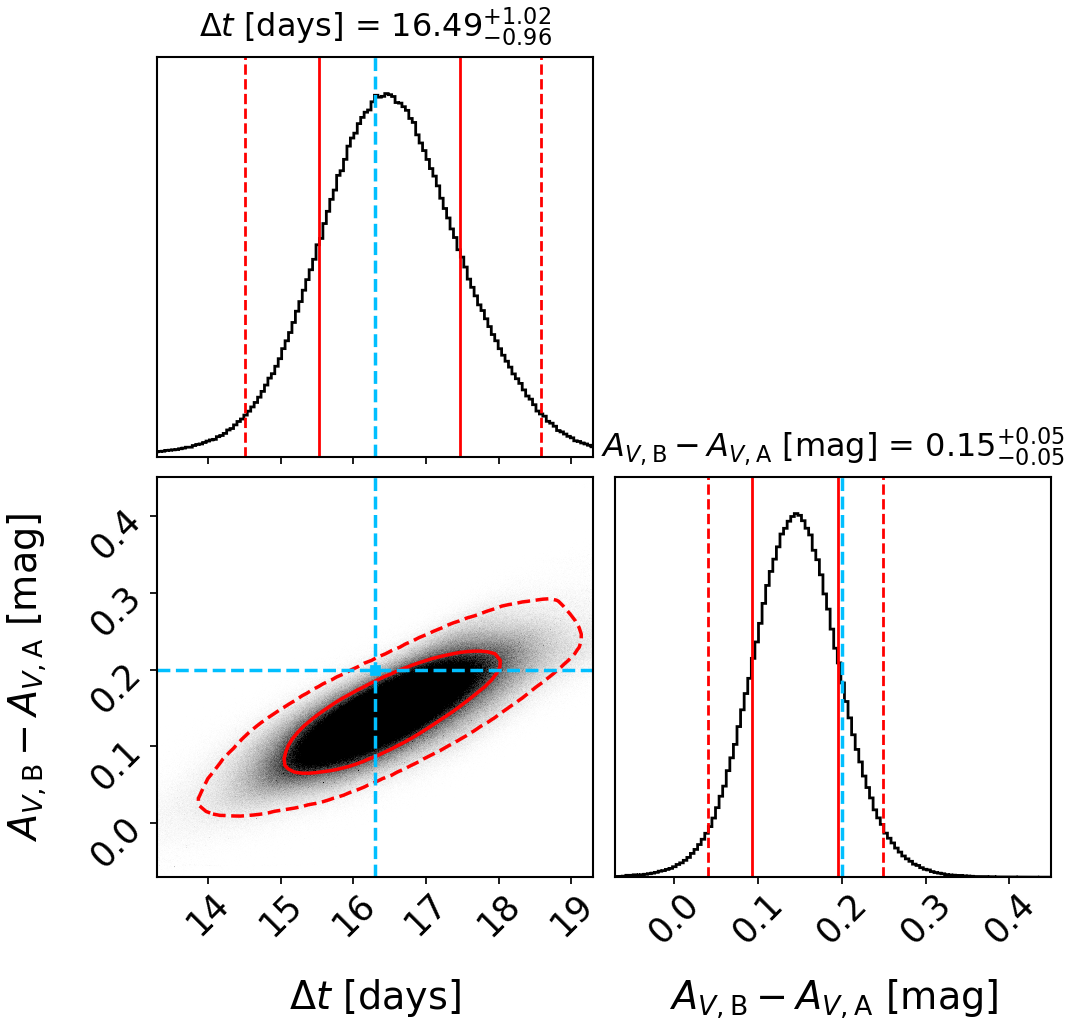}
\caption{\label{RV_2}Corner plot of 100 noise realizations of low S/N mock data with 35\% data loss and with input time delay $\Delta t$ = 16.3 days for combining $u-g$ and $u-r$ color curves setting $R_{V}$ = 2.0 during MCMC sampling.}
\end{figure}

\begin{figure}[hbt!]
\centering
\includegraphics[width=0.49\textwidth]{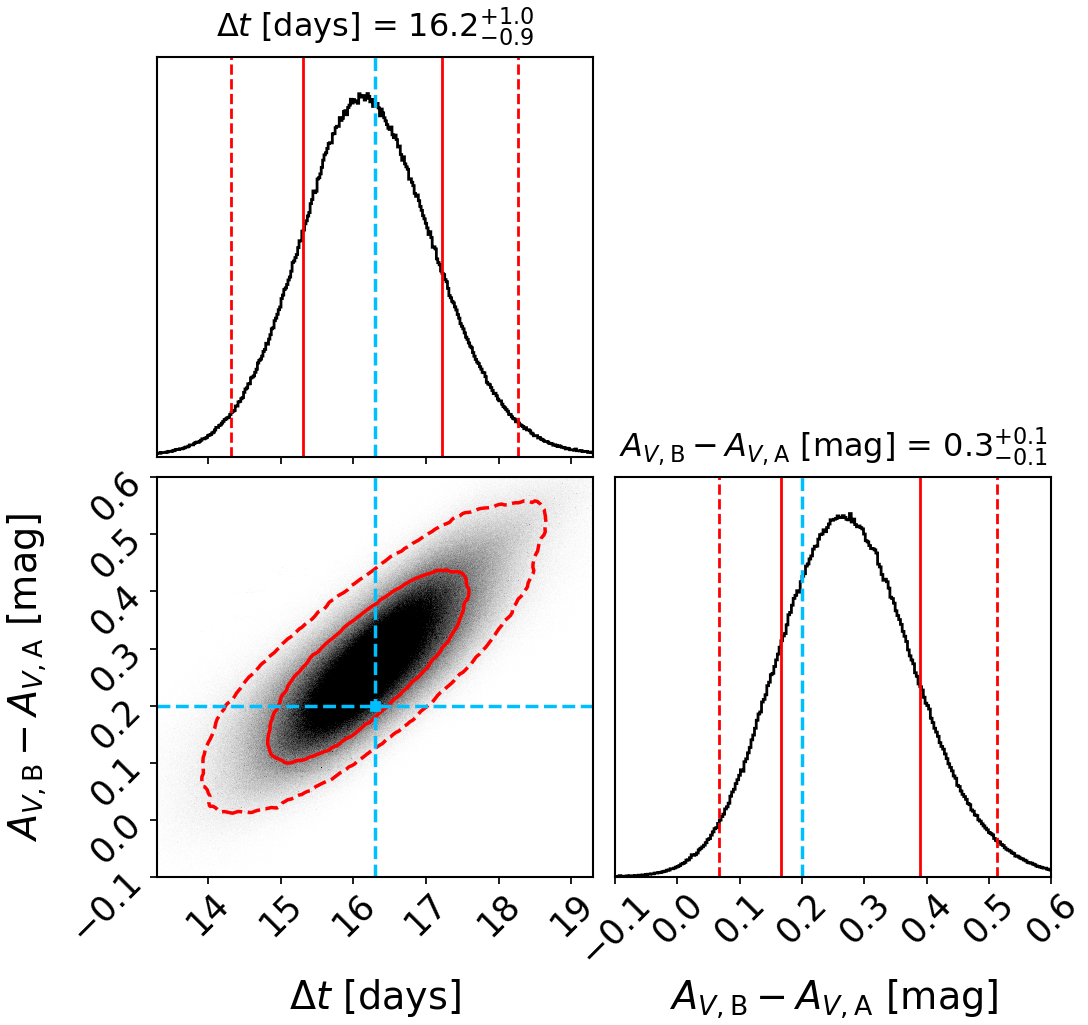}
\caption{\label{RV_5}Corner plot of 100 noise realizations of low S/N mock data with 35\% data loss and with input time delay $\Delta t$ = 16.3 days for combining $u-g$ and $u-r$ color curves setting $R_{V}$ = 5.0 during MCMC sampling.}
\end{figure}

\begin{figure}[hbt!]
\centering
\includegraphics[width=0.49\textwidth]{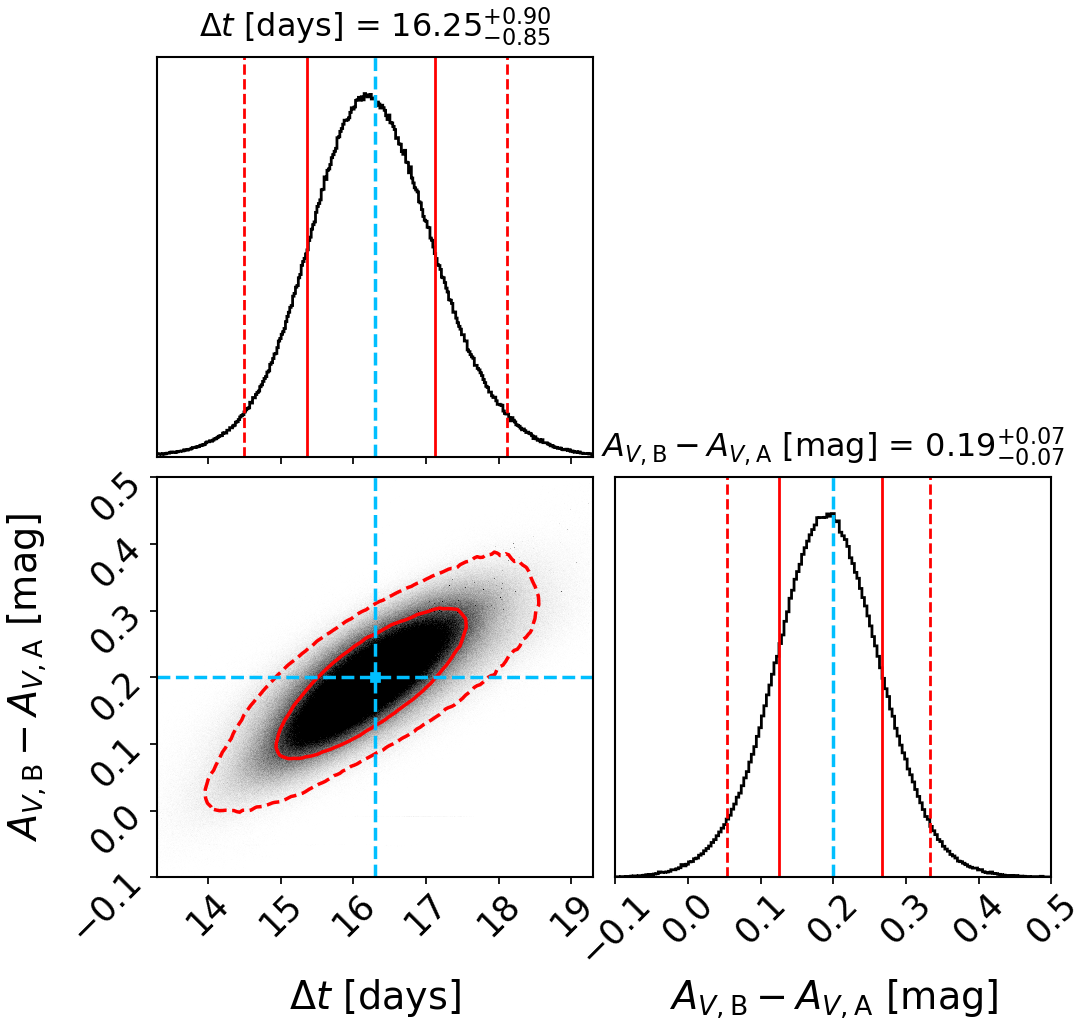}
\caption{\label{RV_Gordon}Corner plot of 100 noise realizations of low S/N mock data with 35\% data loss and with input time delay $\Delta t$ = 16.3 days for combining $u-g$ and $u-r$ color curves using extinction laws by \cite{Gordon2023} with $R_{V}$ = 3.1 during MCMC sampling.}
\end{figure}

The resulting corner plots for T1, T2, and T3 are shown in Fig. \ref{RV_2}, \ref{RV_5}, and \ref{RV_Gordon} respectively. The orthogonal polynomial orders selected via the BIC for all cases follow a similar distribution as shown in Fig. \ref{BIC_ugur_loss}.
For T1 and T2, the time delays have a bias of 0.1-0.2 days, which for delays $>$20 days are still at the $<$1\% level. That is within the tolerance to achieve a 1\% $H_{0}$ measurement.
For short delays, the $R_{V}$ value has a relatively larger impact on the accuracy of the delays, therefore $R_{V}$ should be incorporated in the measurement of the delays.
In both test cases of T1 and T2, $A_{V,\mathrm{B}}-A_{V,\mathrm{A}}$ can not be sampled correctly anymore, which is to be expected considering an incorrect $R_{V}$ value.


For T3 both $\Delta t$ and $A_{V,\mathrm{B}}-A_{V,\mathrm{A}}$ are retrieved unbiased and precisely, because the Cardelli extinction laws and the Gordon laws follow a similar behavior in the considered wavelength range and differ only a bit in the $a(x)$ and $b(x)$ values. If the data would cover the ultraviolet or infrared wavelength range of the extinction laws, the results would deviate more. The same would occur if extinction laws which do not follow the Cardelli laws as closely as the Gordon laws would be applied.

\end{document}